\documentclass[journal]{IEEEtran}

\usepackage{graphicx}  

\usepackage[centertags]{amsmath}   
\begin{document}
%
\title{
\emph{Ex}-house 2D finite-element simulation of the whispering-gallery modes of
axisymmetric electromagnetic resonators
}

%
%
\author{Mark Oxborrow
\thanks{Manuscript received August 23rd, 2006. 
This work was supported by the UK National Measurement System's Quantum Metrology Programme 2004-7. 
M. Oxborrow works at the National Physical Laboratory, Teddington, UK.}}
%
%
%
\markboth{\textit{Ex}-house 2D finite-element simulation of whispering-gallery modes ...}
{Oxborrow
: \emph{Ex}-house 2D finite-element simulation of WG modes ...}
%



\maketitle

\begin{abstract}
It is described, explicitly, how a popular, commercially-available software
package for solving partial-differential-equations (PDEs),
as based on the finite-element method (FEM), can be configured to
calculate the frequencies and fields of the whispering-gallery (WG) modes
of axisymmetric dielectric resonators. The approach is traceable;
it exploits the PDE-solver's ability to accept the definition
of solutions to Maxwell's equations in so-called `\emph{weak form}'.
Associated expressions and methods for estimating
a WG mode's volume, filling factor(s) and, in the case of closed(open)
resonators, its wall(radiation) loss, are provided.
As no transverse approxi-mation
is imposed, the approach remains accurate even for so-called \emph{quasi}-TM
and -TE modes of low, finite azimuthal mode order.
The approach's generality and utility are demonstrated by modeling several non-trivial structures:
(i)~two different optical microcavities  [one toroidal made of silica, the other
an AlGaAs microdisk]; (ii) a 3rd-order microwave Bragg cavity containing alumina layers;
(iii) two different cryogenic sapphire X-band microwave resonators.
By fitting one of the latter to a set of measured resonance frequencies, the
dielectric constants of sapphire at liquid-helium temperature have been estimated.
\end{abstract}
%
\IEEEpeerreviewmaketitle
\section{Introduction\label{sec:Introduction}}
%
%
%
%
%
%
%
%
%
%
\PARstart{E}{xperimental} data are related to physical laws, expressed as equations,
through \emph{models}.
To determine the either fundamental,
phenomenological, or `materials' constants that the model's equations include,
the model must first be solved, and explicitly so, to allow the fitting of
its constants, through (Bayesian) regression, to the experimental data.
The inaccurate solution of a model can sometimes contribute significantly to,
if not wholly dominate, the fitted values' uncertainties.
Improvements in the accuracies of solutions can alone motivate
the (re-)determinations of constants from extant (\emph{i.e.}~`old')
experimental data.
Indeed, the method of solution presented in this paper's section \ref{sec:Method}
is subsequently exploited, in section \ref{sec:PermDet},
to determine the values of certain dielectric constants from the
frequencies at which an electromagnetic resonator was found to
resonate experimentally; here, the `\emph{modeling errors}' dominate
over other uncertainties.
%

Once all relevant physical constants are known to sufficient accuracy, a model's solution can
be exploited in the reverse sense to \emph{simulate} as-yet unrealized experimental embodiments.
Simulation enables the properties of a proposed embodiment to be predicted
and, thus, through modifications, for its performance (with respect to a given application)
to be optimized \emph{without}, or at least at reduced, experimental effort.
Sufficient accuracy in the model's solution itself is, again, vital. %
Though analytical models can adequately treat certain highly symmetrical
structures, the sufficiently accurate solution of less symmetrical (though more practical)
structures typically requires  automated numerical computation --as implemented on a digital computer.
Here, the model's structure must first be represented in some `electronic' format.
Then, the physical equations, often including sets of partial-differential ones, are
encoded and solved for the boundary conditions and constitutive relations that
the structure implies.
Though these two tasks can be implemented by hand-coding in low-level
computer languages, highly developed commercial software packages now exists to
facilitate both: (i) computer-aided-design (CAD) tools and (ii) partial-differential-equation (PDE) solvers,
respectively. Many, though by no means all \cite{kunz93,boriskina03,ctyroky04}, of the latter
are based on the finite-element method (FEM) \cite{zienkiewicz00},
which can readily accept CAD-defined structures.

Furthermore, various packages now integrate (i) and (ii) into complete
computer-modeling environments, \emph{e.g.}~`ECAD' for simulating electromagnetic systems.
These environments sport various additional features for accelerating the definition of models
and for facilitating the display and analysis (`post-processing') of solutions;
they also impose standardized formats and provide (`house-keeping') tools to
assist in the maintenance, sharing and documentation of a model --so that others can
subsequently benefit from, and build upon, the original model-developer's effort.
Compared to the laborious coding up and piecing together of the
equivalent software by hand (\emph{e.g.}~as straight MATLAB
or Fortran code, making calls to optimized `canned' matrix eigensolvers),
the use of such environments, despite their costs and limitations,
is attractive.

A problem associated with the inclusion of a complex model
into the determination of a constant, where the model is solved
via a piece of commercial `black-box' software,
or through a proprietary `in-house' service, is that
the determination may thus cease to be \emph{traceable}.
Significant effects (or `undocumented features')
imparted by the modeling/simulation process may become difficult if not
impossible to isolate, understand, or quantify.
With regard to traceability, both the model's definition and its chosen method of
solution must remain amenable to explicit representation, thus communication,
thus external scrutiny. Convenience and/or efficiency demand, furthermore,
that this representation be as concise and elegant as possible --with no ambiguities.

\subsection{Whispering-gallery-mode resonators\label{subsec:WhisperingGallery}}
Electromagnetic structures that support whispering-gallery (WG) modes  are technologically important because
of the advantageous properties that these modes exhibit in terms of spatial compactness, frequency control
(either stability or agility) and mode quality ($Q$) factor. Explicit examples will be presented
and/or referenced in due course.
Compared to the abrupt retro-reflection of an
electromagnetic wave at the surface of a mirror, the continuous bending of the same by a whispering-gallery
waveguide is an alternative, and at that a still relatively underdeveloped one, which is
opening up new applications. Here, one is often interested in a \emph{`resonator'} where the
wave's trajectory closes back on (and the wave thereby interferes with) itself.
Though elliptical \cite{nockel97}, helical, or even more complex bending trajectories  \cite{sumetsky04}
can be (and have been) envisaged in association with the various morphologies of electromagnetic waveguide/resonator
that support WG waves, the author restricts himself here to the study of the simplest,
and to-date most popular, class of WG-mode resonator: that where the electromagnetic wave's trajectory is
a plane circle (thus constant radius of curvature) and the electromagnetic structure supporting it
is axisymmetric (and coaxial with respect to the said circle/WG mode).\footnote{It is acknowledged that
even axisymmetric (3D) resonators can support `spooling'-helical WG modes that do not lie in
a plane \cite{sumetsky04}; the analysis of such exotica lies outside the scope of this paper.}
Within this class, a convex dielectric:vacuum boundary is often the curved interface of choice
for guiding/confining the whispering gallery mode around in an circle.
The method presented below can,
however, also be employed to simulate WG-modes that are guided by a concave dielectric:metal boundaries.
In general then, one considers an axisymmetric toroidal volume, whose cross-section in a (it matters not which)
radial-axial plane comprises regions of dielectric (voids correspond to the dielectric vacuum) that are bounded (either
externally or from within) by metal surfaces; see FIG.~1.

Despite the breadth and technological allure of this class of WG-mode resonator,
it is the author's understanding that most if not all commercial (ECAD)
packages available at the time of writing (early to mid.~2006) suffer from a rather
unfortunate `blind spot' when it comes to calculating, efficiently (hence accurately),
the whispering-gallery modes (with plane circular trajectories) that such axisymmetric resonator's support.
The popular MAFIA/CST package \cite{mafia}, with which the author is familiar, is a case
in point: As has also been experienced by Basu \emph{et al} \cite{basu04}, and no doubt others,
one simply cannot configure the software to take advantage of the WG modes' \emph{apriori}-known
azimuthal dependence, \emph{viz.}~$\textrm{exp}(\textrm{i} M \phi)$, where $M$ is a positive
integer known as the mode's azimuthal mode order, and $\phi$ is the azimuthal coordinate.
Though frequencies and field-patterns can be obtained (at least for WG modes
of low azimuthal mode order), the computationally advantageous reduction of the problem
from 3D to a 2D that the rotational symmetries of the resonator and its solutions allow is, consequentially, precluded;
and the ability to simulate high-order whispering-gallery modes with sufficient accuracy for metrological purposes
is, exasperatingly, lost.
About the best one can do is to simulate a `wedge' [over an azimuthal domain $\Delta  \phi = \pi/(2 M)$ wide]
between radial electric and magnetic walls \cite{oxborrow01}.

\noindent \emph{From ECAD to `omni-CAD':} Adding titillation to the exasperation,
several commercial packages \cite{COMSOL,ansys}
based on the FEM method are now beginning to offer true `multiphysics' capabilities:
not only can one separately model a resonator's electromagnetic response, its mechanical response,
its thermal response, ..., all based on a common, defined-once-and-for-all geometric structure,
one can furthermore couple/`extrude'/integrate these responses to model non-linear and/or parametric effects.
These effects include (as illustrative examples): (i) the electromagnetic heating of a resonator's lossy dielectrics
and/or it resistively conducting inner surfaces (thus shifts in the frequencies
of the resonator's electromagnetic modes), and --even-- (ii) `mechanical-Kerr'
instabilities/oscillations associated with the mechanical deformation of the resonator's
components due to radiation pressure \cite{rokhsari05}, as exerted by a driven electromagnetic mode.
This \emph{nirvana} of predictive (+ deductive) capability is, for WG-mode resonators,
in view of the alluring applications associated with their nonlinear and/or parametric effects,
a particularly tantalizing destination --if only one could appropriately configure the
(in-the-first-place sufficiently configurable) software to get there.
This paper provides a single, though --one might claim-- a quite fundamental, generic,
and enabling, step on the long march there to.

\subsection{Brief, selected history of WG-mode simulation\label{subsec:BriefHistory}}
The analysis and modeling of the whispering-gallery modes of electromagnetic resonators,
at both optical and microwave frequencies, continues to support and guide experimental
endeavor \cite{spillane05,srinivasan06,wolf04,krupka05}.
A brief and far-from-comprehensive survey of the different methods used to implement
these simulations to date, with a strong selective bias towards those that have been
applied to the study of microwave dielectric-ring resonators, is provided here.
The author encourages the reader to consult the earlier works that are referenced
within the papers cited below.

Based on `separating the variables' (SV), textbooks \cite{ramo84,inan00}
provided expressions for the whispering-gallery modes that are supported by whole
dielectric parallelopipeds, right-cylinders, and spheres, or dielectric layers
and shells exhibiting the same symmetries thereof, where the dielectric volume
is enclosed by electric and/or magnetic walls.
Illustrating the genre is Wilson \emph{et al}'s handy study of the transverse electric (TE)
and transverse-magnetic (TM) modes of right-cylindrical metal cavities \cite{wilson46},
which was in fact used by the author to validate the weak-form expressions
described in sub-section \ref{subsec:WeakForms} below in the early stages of his work.

By `mode-matching' (MM) these SV-solutions across boundaries, WG-mode solutions for composite
axisymmetric structures, such as a dielectric right cylinder, surrounded immediately by a void,
enclosed within a (coaxial) right-cylindrical metallic jacket (\emph{i.e.}, a so-called `can'),
can be derived \cite{tobar91a,tobar04}.
These solutions, with their associated discrete/integer indices (related to symmetries),
provide a nomenclature \cite{krupka99a} for classifying WG modes. This nomenclature
can be re-used to sort and label the lower-order WG modes of less symmetrical though structurally
similar resonators, where these modes can only be calculated `blindly'
through other, more numerical methods.
Mode-matching by taking linear combinations of several/many --as opposed to just a few-- basis functions
can increase the `fit' hence accuracy of the MM-SV method and/or allow it to be extended it to the
treatment of deformed structures.

In view of the limited computational effort that these semi-analytical SV-based methods demand,
remarkable accuracies can be achieved, especially when the most `sympathetic' basis
functions are deployed. For many resonators of interest, however, and the composite axisymmetric
structures mentioned in the previous paragraph are a case in point,
it is not possible, in the MM-SV method (as based on
a \emph{finite} set of basis functions), to simultaneously match all components of the electric and
magnetic fields across all boundaries \cite{wolf04} --to do so would, after all, amount to an exact
solution! With a small, finite basis set, the `transverse' (or `axial-polarization') approximation,
that tolerates a mis-match of  `minor' field components, whilst consistently matching the major ones,
is uncontrolled. Though extensions to mode-matching that capture
spatially non-uniform polarization can be constructed \cite{ctyroky04}, the MM-SV method
in general needs to be validated, for a given shape of resonator and mode, through comparison
with (more exact) solutions supplied by other methods.

For a complete, accurate solution of Maxwell's equations, one must generally
resort to wholly numerical methods, of which there are several relevant classes and variants.
Apart from the finite-element method (FEM) itself (considered in more detail further on),
the most developed and thus most immediately exploitable alternatives include
(given here for reference --not considered in any greater detail):
(i) the Ritz-Rayleigh variational or `moment' methods \cite{krupka94,krupka96,monsoriu02},
(ii) the finite difference time domain method (FDTD) \cite{kunz93,alford00}, and (iii)
the boundary-integral \cite{boriskina03} or bounday-element methods (BEM, including FEM-based hybridizations
thereof \cite{wolf03}).
Zienkiewicz and Taylor \cite{zienkiewicz00}, though nominally dedicated to FEM,
provide a taxonomy (\emph{viz.}~table 3.2 \emph{loc. cit.}) covering most of these methods,
which reveals certain commonalities between them. It is remarked here, for example, that
FDTD may be regarded as a variant of the FEM employing local, discontinuous shape functions.\footnote{It
is also worth remarking that, for resonators comprising just a few, large domains
of uniform dielectric, then the boundary-integral methods (based on Green functions),
which --in a nutshell-- exploit such uniformity to reduce the problem's dimensionality by one,
will generally be more computationally efficient than FEM.} In conjunction with these generalities,
it is worth re-iterating here that
the core formulation presented in this paper (\emph{viz.}~equations
\ref{eq:laplacianweakall} through \ref{eq:magneticwallHcylcomp2})
can exploit any PDE-solver (\emph{e.g.}~a moment-method-based one) capable of
accepting/intepreting weak-form statements.
Though a FEM-based solver (\emph{viz.}~COMSOL/FEMLAB) was indeed used to
provide the examples presented in section \ref{sec:ExampleApplications} below,
this article's formulation is not, \emph{per se}, wedded to FEM.

Though the finite-element method can solve for all field components (both major and minor),
the explicit, direct statement of the required set of a coupled partial differential equations
(\emph{i.e.}~Maxwell's equations for WG-mode electromagnetic resonators) in component form,
suitable for the insertion into a standard commercial FEM/ECAD software package, can be extremely
onerous --if not absolutely ruled out by the software's lack of configurability.
This is why the majority of these packages already include pre-defined `applications modes',
`macros' or `wizards' for solving electromagnetic problems (for particular geometries).
To simplify the problem to that of a single (scalar) PDE, one can again [cf.~the SV-MM method(s)
already discussed above] invoke the so-called transverse (axial-polarization) approximation,
where the resonator's either magnetic or electric
field is forced to lie everywhere parallel to the resonator's axis of rotational symmetry;
figure B.1 of reference \cite{kippenberg04} displays this approximation most pedagogically.
Investigations based on this `transverse-FEM' approach have been reported in several recent
works \cite{basu04,kippenberg04}.\footnote{Srinivasan \emph{et al} \cite{srinivasan06} in contrast
employ a `full-vectorial model' though they do not explicitly define what this is.}
Though it can provide indicative trends and quantitative results, which might well be
sufficiently accurate and/or robust for the calculation's intended purpose (in view of even less well
controlled experimental parameters), the uncontrolled nature of the approximation that transverse-FEM
incorporates is again far from ideal.
The careful physicist, or metrologist, is (again) compelled to justify its validity,
for a given resonator and mode, through comparison with either non-trivial analytical
solutions, where they exist, or `brute force' (3D) numerical computation \cite{basu04}.
It is this paper's principal claim that, through only a modicum of extra effort,
the transverse approximation, and its associated onerous validations [or (else) lingering doubts],
can be wholly obviated.

The application of the finite-element method to the solving of Maxwell's equations
has a history \cite{rahman91}, and is now very much an industry \cite{mafia,COMSOL,ansys}.
Zienkiewicz and Taylor \cite{zienkiewicz00} supply FEM's theoretical
underpinnings, in particular an erudite account of Galerkin's method of
so-called `weighted residuals'.
A pervasive, and often quite debilitating problem that besets the direct/`na\"{i}ve' applications of
FEM to the PDEs that are Maxwell's equations is the generation of (a great many) spurious solutions \cite{auborg91,lee93},
associated with the local gauge invariance, or `null space' \cite{lee93}, which is a (hidden) feature
of these PDEs (in particular its `curl' operators).
At least two research groups have nevertheless successfully developed software tools for
calculating the WG modes of axisymmetric dielectric resonators, where these tools
(i) solve for all field components (\emph{i.e.}~no transverse approximation is invoked),
(ii) are 2D (and thus numerically efficient) and (iii)
effectively suppress spurious solutions (without any insidious, detrimental
side-effects) \cite{auborg91,krupka94,krupka96,osegueda94,santiago94}.
The method reported below sports these same three attributes.
With regard to (iii), the approach adopted by Auborg \emph{et al}
was to use different finite elements (\emph{viz.} a mixture of  `Nedelec' and `Lagrange' -both 2nd order)
for different components of the electric and magnetic fields;
Osegueda \emph{et al} \cite{osegueda94}, on the other hand, used a so-called
`penalty term' to suppress (spurious) divergence of the magnetic field.
Stripping away all of its motivating remarks, applications and illustrations, the
principal function of this paper is to convert (one might say `extract') the method
encoded by Osegueda \emph{et al}'s `CYRES' 2D FEM software
package \cite{osegueda94} into explicit \emph{`weak-form'} expressions,
that can be directly and \emph{openly} ported to \emph{any} PDE solver (most notably COMSOL/FEMLAB) capable
of accepting such. These weak-form expressions are wholly equivalent to the Maxwellian PDEs from which they are
derived. But, being scalar (tensorially-contracted), they are considerably less onerous to represent and
communicate than the vectorial PDEs themselves. The  author hopes that, by stating/popularizing the problem so
explicitly in this paper through the \emph{lingua franca} of weak form, the means to model, both accurately
and traceably, the whispering-gallery modes of axisymmetric resonators will thereby be made accessible
to any competent engineer or physicist in need of such a means --`off the shelf', as opposed
to it remaining a strictly `\emph{in}-house' (and thus rather less open and traceable)
capability retained by specialists.

\section{Method of solution\label{sec:Method}}

\subsection{Weak forms\label{subsec:WeakForms}}
\begin{figure}[h]
\centering
\includegraphics[width=0.75\columnwidth]{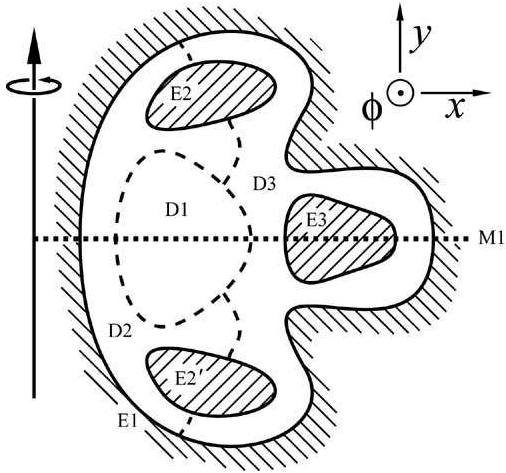}
\caption{Generic axisymmetric resonator in cross-section (medial half-plane). A dielectric
region enclosed by an electric wall E1 is subdivided into subdomains D1, D2 and D3, none, one or more
of which could be free space. D2 and D3 are bounded internally by electric walls E2,
E2$'$ and E3. The resonator's mirror symmetry through the horizontal (equatorial)
plane, as indicated by the dashed line M1, allows an additional either electric or
magnetic wall to be placed on it and, thereupon, only one half of the resonator's domain
(either its upper or lower half) need be simulated.}
\label{fig:generic_resonator}
\end{figure}

\noindent \emph{Scope:}
The types of ideal resonator that fall under the scope of the analysis presented immediately below
are those that comprise volumes of lossless dielectric space bounded by a combination of perfect
(thus also loss-less) electric or magnetic walls --see Fig.~\ref{fig:generic_resonator}
(though note that the restriction to resonators of axisymmetric form needs only to be invoked
at the start of subsection \ref{subsec:Axisymmetric}).
As discussed in subsections \ref{subsec:WallLosses} through \ref{subsec:RadiationEstimators} below,
a dissipative (open) resonator's finite fractional energy loss per cycle, hence its sub-infinite $Q$-factor,
can be estimated, and with often perfectly sufficient accuracy with regard to applications,
from the solution of an equivalent loss-less (closed) resonator.
The resonator's dielectric space will, in general, comprise both voids (\emph{i.e.}~the free space
of a vacuum) and space filled by sufficiently `good' (\emph{i.e.}~low-loss) dielectric material(s).
A model's electric walls will translate, in embodiments, to metallic surfaces whose conductivities are sufficiently
good to be treated as such (section \ref{subsec:WallLosses} quantifies the loss caused
by the metallic wall of a particular resonator's can).
The (relative) permeability of real magnetic materials is seldom high enough for a wall made
from such to be regarded, without deleterious approximation, as a (perfect) magnetic one.
When modeling resonators whose forms exhibit reflection symmetries, such that the
magnetic and electric fields of their solutions exhibit either symmetry or antisymmetry
through mirror planes, perfect magnetic and electric walls can be advantageously imposed
on the model's mirror plane to solve for particular `sectors' of solutions.

The electromagnetic field within the dielectric volumes of the resonator obeys Maxwell's
equations \cite{bleaney76,inan00}, as they are applied to continuous, macroscopic media \cite{robinson73}.
Thus, on the assumption that the resonator's constituent dielectric elements have negligible (or at
least the same) magnetic susceptibility (hence permeability), the magnetic field strength $\textbf{H}$
is continuous across all interfaces between them.\footnote{The method described in this paper could be
extended (still within the configurability of the PDE-solver) to the analysis of resonators containing
different magnetic dielectrics (such as YIG, and as would exhibit differing permeabilities) by way of `coupling variables'
--to transform $\textbf{H}$ (or $\textbf{B}$) across internal boundaries between regions of different permeability.}
This property makes it advantageous to solve for $\textbf{H}$ (or the magnetic flux
density $\textbf{B} = \mu \textbf{H}$ related to it by a constant global magnetic permeability $\mu$),
as opposed to the electric field strength $\textbf{E}$ (or displacement $\textbf{D}$).
Upon substituting one of Maxwell's `curl' equations equations into the another,
the problem reduces to that of solving the (modified) vector Helmholtz equation
\begin{equation}
\label{eq:helmholtz}
\boldsymbol{\nabla} \pmb{\times} ({\boldsymbol{\epsilon}^{-1}}\; \boldsymbol{\nabla} \pmb{\times} \textbf{H})
-\alpha \boldsymbol{\nabla} (\boldsymbol{\nabla} \cdot \textbf{H})
+ c^{-2} \partial^2\textbf{H}/{\partial t^2} = 0
\end{equation}
subject to appropriate boundary conditions (read on), where $c$ is the speed of light.
Here, ${\boldsymbol{\epsilon}^{-1}}$ denotes the inverse relative permittivity tensor;
one assumes that the resonator's dielectric elements are linear, such that it is
a (tensorial) constant --\emph{i.e.}~independent of field strength.
The middle (penalty) term on the left-hand side of
equation \ref{eq:helmholtz} is the same as that used by
Osegueda \emph{et al} \cite{osegueda94}; it functions to
suppress/reveal spurious modes in the finite-element simulation\footnote{It is mentioned here that the author briefly experimented in COMSOL
with mixed (`Nedelec' plus `Lagrange') finite elements \cite{krupka94} but found
that the above penalty term, in conjunction with 2nd-order Lagrange finite
elements applied to all three components of $\textbf{H}$, gave wholly satisfactory results.};
the constant $\alpha$
controls this term's weight with respect to its Maxwellian neigbours.
The penalty term's insertion into the above Helmholtz equation
is wholly permissible since, for every true solution of Maxwell's equation, it must exactly
vanish (everywhere): the magnetic flux density $\textbf{B}$,
hence (for non- or isotropically-magnetic media)
$\textbf{H} = \textbf{B}/\mu$, is required to be free of divergence
--assuming no magnetic monopoles lurk inside the resonator.

Reference~\cite{inan00} (particular section 1.3 thereof) supplies a primer
on the electromagnetic boundary conditions discussed forthwith.
Assuming that the resonator's bounding electric walls are perfect in the sense of
having (effectively) infinite surface conductivity, the magnetic flux density
at any point on each such wall is required to satisfy $\textbf{B} \cdot \textbf{n} = 0$,
where $\textbf{n}$ denotes the wall's surface normal vector.
Providing the magnetic permeability/susceptibility of the dielectric medium
bounded by the electric wall is not anisotropic, this condition is equivalent to
\begin{equation}
\textbf{H} \cdot \textbf{n} = 0.
\label{eq:electricwallH}
\end{equation}
The electric field strength at the electric wall is required to obey
\begin{equation}
\textbf{E} \pmb{\times} \textbf{n} = 0;
\label{eq:electricwallD}
\end{equation}
these two equations ensure that the magnetic (electric) field strength
is oriented tangential (normal) to the electric wall. As is pointed out
in reference~\cite{osegueda94}, equation \ref{eq:electricwallD} is a so-called
`natural' (or, synonymously, a `naturally satisfied') boundary condition
within the finite-element method --see ref.~\cite{zienkiewicz00}.

When the resonator's form (hence those of its solutions) exhibits one or more
symmetries, it is often advantageous (for reasons of computational efficiency) to
solve only for a symmetry-reduced portion or `sector' of the full resonator,
where this sector is bounded by
either (real or virtual) electric walls or (virtual) magnetic walls, or both.
The boundary conditions corresponding to a perfect magnetic wall
(dual to the those for an electric wall) are
\begin{equation}
\label{eq:magneticwallD}
\textbf{D} \cdot \textbf{n} = 0,
\end{equation}
and
\begin{equation}
\label{eq:magneticwallH}
\textbf{H} \pmb{\times} \textbf{n} = 0;
\end{equation}
these two equations ensure that the electric displacement (magnetic field strength)
is oriented tangential (normal) to the magnetic wall. Again, the latter
equation is naturally satisfied.

One now invokes Galerkin's method of weighted residuals; reference \cite{zienkiewicz00} explains
the fundamentals here; reference \cite{lee93} provides an analogous treatment when solving for
the electric field strength ($\textbf{E}$). Both sides of equation \ref{eq:helmholtz} are multiplied
(scalar-product contraction) by the complex conjugate of a `test' magnetic field strength $\tilde{\textbf{H}}^*$,
then integrated over the dielectric resonator's interior volumes. Upon expanding the permittivity-modified
`curl of a curl' operator (to extract a similarly modified Laplacian operator), then integrating by
parts (spatially), then disposing of surface terms through the electric- or magnetic-wall
boundary conditions stated above, one arrives (equivalent to equation (2) of reference \cite{osegueda94}) at
\begin{multline}
\label{eq:weakhelm}
\int_{\textrm{V}} [ (\boldsymbol{\nabla} \pmb{\times} \tilde{\textbf{H}}^*) \; \frac{\cdot}{\boldsymbol{\epsilon}} \;
(\boldsymbol{\nabla} \pmb{\times} \textbf{H}) -
~~~~~~~~~~\\
\; \alpha (\boldsymbol{\nabla} \cdot \tilde{\textbf{H}}^*) (\boldsymbol{\nabla} \cdot \textbf{H}) +
c^{-2} \tilde{\textbf{H}}^* \cdot \partial^2\textbf{H}/{\partial t^2} ] \, \textrm{d} \textrm{V} = 0,
\end{multline}
where `$\int_{\textrm{V}}$' denotes the volume integral over the resonator's interior space (or sector thereof)
and `$\cdot/\boldsymbol{\epsilon}$' denotes a contraction weighted by inverse relative permittivities.
The three terms appearing in the integrand correspond directly to the three weak-form terms required to define an
appropriate finite-element model within the PDE solver.

Assuming that the physical dimensions and electromagnetic properties of the resonator's
components are temporally invariant (or at least `quasi-static'), one
seeks harmonic or `modal' solutions:
$\textbf{H}(\textbf{r};t) = \textbf{H}(\textbf{r}) \textrm{exp}(-\textrm{i} 2 \pi f t)$,
where $\textbf{r}$ is the vector of spatial position,
$t$ the time, and $f$ the mode's resonance frequency.
The last, `temporal' term in equation \ref{eq:weakhelm}'s integrand can thereupon be re-expressed as
$-(\bar{c} f)^2 \tilde{\textbf{H}}(\textbf{r})^* \cdot \textbf{H}(\textbf{r})$,
where $\bar{c} \equiv 2 \pi/c$ and $c$ is the speed of light. This re-expression(and, with respect to Spillane \emph{et al}'s work, using exactly the same FEM software platform)
reveals the integrand's complete dual symmetry between $\tilde{\textbf{H}}^*$ and
$\textbf{H}$.

\subsection{Axisymmetric resonators\label{subsec:Axisymmetric}}
One now restricts the scope of the analysis to resonators whose interiors
and bounding surfaces are electromagnetically axisymmetric (henceforth referred to simply as
`axisymmetric resonators') where a system of cylindrical coordinates is
aligned with respect to the resonator's axis of rotational symmetry.
This system's three components are
$\{x, \phi, y\} \equiv$ \{`rad(ial)', `azi(muthal)', `axi(ial)'\}\footnote{`$x$'
and `$y$', instead of the more conventional `$r$' and `$z$',
are (regrettably) used to represent radial and axial coordinates/components,
respectively, so as to comply with COMSOL/FEMLAB's standard (2D) naming conventions.}.
One wishes to calculate the resonance frequencies and field patterns of the
resonator's (standard) whispering-gallery (WG) modes, whose phase varies as
$\textrm{exp}(\textrm{i} M \phi)$, where $M =\{0, 1, 2, ...\}$ is the WG mode's
azimuthal mode order. Note that the method does not require $M$ to be large
(\emph{i.e.}, it is not an `asymptotic' method);
even modal solutions that are themselves axisymmetric, corresponding to $M=0$,
such as the one shown in Fig.~\ref{fig:BraggCavity}(b), can be calculated.
Viewed as a three-component vector field over a (for the moment) three-dimensional space,
the time-independent part of the magnetic field strength now takes the form
\begin{equation}
\label{eq:Hcylin}
\textbf{H}(\textbf{r}) =  \textrm{e}^{\textrm{i} M \phi} \left\{ \right.  H^x(x,y),
\textrm{i} \, H^\phi(x,y),
H^y(x,y) \left. \right\}
\end{equation}
where an `$\textrm{i}$' ($ \equiv \surd(-1)$) has been inserted into the field's azimuthal component
so as to allow, in subsequent solutions, all three component amplitudes
$\left\{ H^x, H^\phi, H^y \right\}$ to each be expressible as a real amplitude
multiplied by a common complex phase factor.
The relative permittivity tensor of an axisymmetric dielectric material is diagonal
with entries (running down the diagonal)
$\boldsymbol{\epsilon}_\textrm{diag.} = \{\epsilon_\perp, \epsilon_\perp, \epsilon_\parallel\}$,
where $\epsilon_\parallel$ is the material's relative permittivity in the axial direction
and $\epsilon_\perp$ its relative permittivity in the plane spanned by it radial and azimuthal directions.

It now remains only to substitute equation \ref{eq:Hcylin} into equation \ref{eq:weakhelm}'s integrand
and express the three terms composing the latter's integrand in terms of the magnetic field strength's components
(and their spatial/temporal derivatives); textbooks provide the required explicit expressions for
the vector differential operators in cylindrical coordinates \cite{marder70,ramo84,inan00}.
A radial factor, $x$, is included here from the volume
integral's measure: $\textrm{d} \textrm{V} = 2 \pi \, x \, \textrm{d} x \, \textrm{d} \phi \, \textrm{d} y$ (the
factor of $2 \pi$ here is uniformly, thus inconsequentially, dropped from all three
expressions below.)
These requisite expansions are presented here in compact mathematically notation;
their line-text (\emph{i.e.~}with no super- or sub-scripts, hence considerably more verbose)
equivalents, in forms suitable for
direct cut-and-paste injection into a popular PDE-solver (\emph{viz.} COMSOL/FEMLAB)
are available as a separate `Appendix' to this paper \cite{oxborrow06}.
The first, `Laplacian' weak term is given by
\begin{equation}
\label{eq:laplacianweakall}
(\boldsymbol{\nabla} \pmb{\times} \tilde{\textbf{H}}^*) \frac{\cdot}{\boldsymbol{\epsilon}} \;
(\boldsymbol{\nabla} \pmb{\times} \textbf{H}) =
\bigl( \frac{A}{x} + B + x \,C \bigr) / ({\epsilon_\perp \epsilon_\parallel}),
\end{equation}
where
\begin{multline}
\label{eq:laplacianweakA}
A \equiv \{\,\epsilon_\perp [\tilde{H}^\phi H^\phi
- M (\tilde{H}^\phi H^x + H^\phi \tilde{H}^x  ) + M^2 \tilde{H}^x H^x )] \\
~~~~~~+\epsilon_\parallel M^2 \tilde{H}^y H^y  \, \},
\end{multline}
\begin{multline}
\label{eq:laplacianweakB}
B \equiv \epsilon_\perp [ \tilde{H}^\phi_x ( H^\phi - M H^x ) + H^\phi_x (\tilde{H}^\phi - M \tilde{H}^x)]\\
~~~~~-\epsilon_\parallel M (\tilde{H}^y H^\phi_y  + H^y \tilde{H}^\phi_y  ),
\end{multline}
\begin{multline}
\label{eq:laplacianweakC}
C \equiv \{ \,\epsilon_\perp \tilde{H}^\phi_x H^\phi_x \\
~~+\epsilon_\parallel [(\tilde{H}^y_x - \tilde{H}^x_y) (H^y_x-H^x_y) + \tilde{H}^\phi_y H^\phi_y ] \, \},
\end{multline}
where $H^\phi_x$ denotes the partial derivative of $H^\phi$ (the aziumthal component of the magnetic field strength)
with respect to $x$ (the radial component of displacement), \emph{etc.}.
Here, the individual factors and terms have been ordered and grouped
so as to display the dual symmetry. Similarly, the weak penalty term is given by
\begin{equation}
\label{eq:penaltyweakall}
\alpha (\boldsymbol{\nabla} \cdot \tilde{\textbf{H}}^*) (\boldsymbol{\nabla} \cdot \textbf{H})
= \alpha \{ \frac{D}{x} + E + x \,F \},
\end{equation}
where
\begin{multline}
\label{eq:penaltyweakD}
D \equiv \tilde{H}^x H^x  - M ( \tilde{H}^\phi H^x + H^\phi\tilde{H}^x )+ M^2 \tilde{H}^\phi H^\phi,
\end{multline}
\begin{multline}
\label{eq:penaltyweakE}
E \equiv  (\tilde{H}^x_x + \tilde{H}^y_y) (H^x - M H^\phi)\\
+ (\tilde{H}^x - M \tilde{H}^\phi) (H^x_x + H^y_y), ~~~~~~~~~~~
\end{multline}
\begin{equation}
\label{eq:penaltyweakF}
F \equiv ( \tilde{H}^x_x + \tilde{H}^y_y) (H^x_x + H^y_y).~~~~~~~~~~~~~~~~~~~~~~~~~~~
\end{equation}
And, finally, the temporal weak-form (`dweak') term is given by
\begin{multline}
\label{eq:temporalweak}
\tilde{\textbf{H}}^* \cdot \partial^2\textbf{H}/{\partial^2 t} =
c^{-2} \, x \, ( \tilde{H}^x  H^x_{tt}  + \tilde{H}^\phi H^\phi_{tt} + \tilde{H}^y H^y_{tt}) ~~~~~~~~~~~~~~~~~~~\\
~~~~~~~~~ = -\bar{c}^2 f^2 \, x \, ( \tilde{H}^x  H^x + \tilde{H}^\phi H^\phi + \tilde{H}^y H^y),
\end{multline}
where $H^x_{tt}$ denotes the double partial derivative of $H^x$ w.r.t.~time, \emph{etc.}.
What is crucial is that none of the terms on the right-hand sides of equations
\ref{eq:laplacianweakall} through \ref{eq:temporalweak} depend on the azimuthal coordinate $\phi$;
the problem has been reduced from 3D to 2D.

\subsection{Axisymmetric boundary conditions\label{subsec:AxisymmetricBCs}}
An axisymmetric boundary's unit normal in cylindrical components can
be expressed as $\{n_x, 0, n_y\}$ --note vanishing azimuthal component.
The full electric-wall boundary conditions, in cylindrical components, are as follows:
$\textbf{H} \cdot \textbf{n} = 0$ gives
\begin{equation}
\label{eq:electricwallHcylcomp}
H^x n_x + H^y n_y = 0,
\end{equation}
and $\textbf{E} \pmb{\times} \textbf{n} = 0$ includes both
\begin{equation}
\label{eq:electricwallEcylcomp1}
H^x_y - H^y_x = 0
\end{equation}
and
\begin{equation}
\label{eq:electricwallEcylcomp2}
\epsilon_\perp (H^\phi -H^x M + H^\phi_x x) n_x - \epsilon_\parallel(H^y M - H^\phi_y x) n_y = 0.
\end{equation}
When the dielectric permittivity of the medium bounded by the electric wall is isotropic
(which is often the case in embodiments), the last condition reduces to
\begin{equation}
\label{eq:electricwallEcylcomp2isotrop}
(H^\phi -H^x M + H^\phi_x x) n_x - (H^y M - H^\phi_y x) n_y = 0.
\end{equation}
%
%
%
%
%
%
%
The full magnetic-wall boundary conditions, in cylindrical components,
are as follows: the condition $\textbf{D} \cdot \textbf{n} = 0$ gives
\begin{equation}
\label{eq:magneticwallDcylcomp}
(H^y M - H^\phi_y x) n_x + (H^\phi -H^x M + H^\phi_x x) n_y = 0,
\end{equation}
and the condition $\textbf{H} \pmb{\times} \textbf{n} = 0$ includes both
\begin{equation}
\label{eq:magneticwallHcylcomp1}
H^y n_x -H^x n_y = 0
\end{equation}
and
\begin{equation}
\label{eq:magneticwallHcylcomp2}
H^\phi =0.
\end{equation}
One observes that the transformation
$\{ n_x \rightarrow - n_y, n_y \rightarrow n_x\}$,
connects equations \ref{eq:electricwallHcylcomp} and \ref{eq:magneticwallHcylcomp1},
and equations \ref{eq:electricwallEcylcomp2isotrop} with \ref{eq:magneticwallDcylcomp}.
Explicit PDE-solver-ready equivalents of \ref{eq:electricwallHcylcomp} through \ref{eq:magneticwallHcylcomp2}
are stated in this paper's auxiliary Appendix \cite{oxborrow06}.
The above weak-form expressions and boundary conditions, \emph{viz.}~equations \ref{eq:laplacianweakall}
through \ref{eq:magneticwallHcylcomp2} are the key
enabling results of this paper: once inserted into a PDE-solver, the WG modes of axisymmetric dielectric
resonators can readily be calculated, as is demonstrated for
particular embodiments in section \ref{sec:ExampleApplications} below.

\section{Postprocessing of solutions\label{sec:Postprocessing}}
Having determined, for each mode, its frequency and all three components of
its magnetic field strength $\textbf{H}$ as functions of position,
other relevant fields and parameters can be derived from them.

\subsection{Remaining Maxwellian fields\label{subsec:Maxwellian}}
Straightaway, the magnetic flux density $\textbf{B} = \textbf{H}/ \mu$; here,
as stated in subsection \ref{subsec:WeakForms} above --but see also footnote 4,
the magnetic permeability $\mu$ is assumed to be a scalar constant, independent of position.
[And for each of the resonators considered in section \ref{sec:ExampleApplications},
$\mu =  \mu_0$ everywhere --to an adequate approximation.]
As no real (`\emph{non}-displacement') current flows within a dielectric,
$\nabla \pmb{\times} \textbf{H}(t) = \partial \textbf{D}(t) / \partial t $,
thus $\textbf{D} = -\textrm{i} (2 \pi f)^{-1} \nabla \pmb{\times} \textbf{H}(t)$.
And, $\textbf{E} = \boldsymbol{\epsilon}^{-1}\textbf{D}$, where
$\boldsymbol{\epsilon}^{-1}$ is the (diagonal) inverse permittivity tensor,
as already discussed in connection with equation \ref{eq:weakhelm} above.

\subsection{Mode volume and filling factor\label{sec:ModeVolume}}
Accepting various caveats (most fundamentally, the problem of mode-volume divergence --see footnote 7;
and inconsistent definitions between different authors ...) as
addressed by Kippenberg \cite{kippenberg04}, the volume of a mode in a dielectric resonator
is here defined as \cite{srinivasan06}
\begin{equation}
\label{eq:mode_volume}
V_{\rm{mode}} = \frac{\int \int \int_{\rm{h.-s.}} \epsilon |\textbf{E}|^2 \textrm{d} \textrm{V} }
{ {\rm{max}} [\epsilon |\textbf{E}|^2 ]},
\end{equation}
where ${\rm{max}} [ ... ]$, denotes the maximum value of its functional argument,
and $\int \int \int_{\rm{h.-s.}} ... \textrm{d} \textrm{V} $ denotes integration over and around
the mode's `bright spot' --where its electromagnetic field energy is concentrated.

\subsection{Filling factor\label{sec:Fillingfactor}}
The resonator's electric filling factor, for a given mode, a given dielectric piece/material, $\rm{diel.}$,
and a given field direction, ($\rm{dir.} \in \{\rm{radial}, \rm{azimuthal}, \rm{axial}\}$), is defined as
\begin{equation}
\label{eq:filling_factor}
F_{\rm{diel.}}^{\rm{dir.}} = \frac{\int \int \int_{\rm{diel.}} \epsilon_{\rm{pol.}} |E^{\rm{dir.}}|^2 \textrm{d} \textrm{V} }
{\int \int \int \epsilon |\textbf{E}|^2 \textrm{d} \textrm{V} },
\end{equation}
where $\int \int \int_{\rm{diel.}} ... \textrm{d} \textrm{V} $ denotes integration
only over those domains composed of the dielectric in question and
$\rm{pol.}$ = $\{\perp,\parallel\}$ for $\rm{dir.}$ = $\{\rm{radial} \; \rm{or} \; \rm{azimuthal}, \rm{axial}\}$.
The numerators and denominators of equations \ref{eq:mode_volume} and \ref{eq:filling_factor}
can be readily evaluated using the PDE-solver's post-processing features.

\subsection{Finite Qs and wall losses \label{subsec:WallLosses}}
So far, the model resonator as per Fig.~\ref{fig:generic_resonator} has been treated as a wholly
loss-less one: its modal solutions have infinite $Q$s or, equivalently, the (otherwise complex)
frequencies of these solutions
are purely real or, equivalently, the solutions' oscillatory electromagnetic fields persist indefinitely.
No energy is dissipated by the dielectric material(s) included within the
resonator (their dielectric loss tangents are presumed to be zero); none is lost through radiation --since the
resonator's bounding perfect electric/magnetic walls allow none to escape; and, being perfect, the current
induced within the wall causes no resistive dissipation.

\emph{Real} resonators, on the other hand, are subject to one or several dissipative processes, \emph{i.e.}~`losses',
that render the $Q$s of resonances finite. This subsection
provides an expression for the rate of a resonator's `wall loss'; section \ref{sec:RadLoss}
goes on to provide bounds on the rate of an open resonator's `radiation loss'. Such estimates
are important since $Q$ values are directly measurable in experiments and, furthermore,
often determine viability and/or performance in applications.
The approach taken here is to build upon (via perturbation theory
and/or `induction') the loss-less model, as it has already been formulated
in section \ref{sec:Method} --as opposed to constructing the additional machinery
required to model resonators
with lossy materials either placed within or clad about them.

\emph{Preliminaries:}
The space that a WG-mode occupies can be broadly divided into three regions: (i) the mode's `near-field', which
includes its bright spot(s) --where the modes e.m. energy density is greatest, (ii) an `evanescent zone', lying
around the near-field, where the mode's field amplitudes (and energy density) decay exponentially with
distance $r$ away from its bright spot and (iii) a (notionally infinite) `radiation zone', lying outside
of the evanescent zone (a `cusp' can separate the two), where the field-amplitudes decay as $\sim 1/r$.
If a compact, closed resonator of high-$Q$ is sought, its electric (\emph{i.e.}~metal) wall should be placed
sufficiently far from the WG-mode's bright spot in the exponentially decaying zone (ii), but,
for reasons of compactness, no further out than is necessary.

A brief word of warning: As the $Q$ of an experimental WG-resonator can be exceedingly high ($> 10^9$),
the amplitude of the electromagnetic field where dominant losses occur (their rates will depend on the amplitude)
can be many orders of magnitude lower than the field's maximum (or maxima) at the WG-mode's bright spot(s). The
hardware and software employed to generate WG-modal solutions must thus be able to cope with such a dynamic
range, lest significant numerical errors creep into the predicted (loss-rate-determining) amplitudes where
the WG-resonator's supported mode is faint.
In practice, this means adequate-precision arithmetic, adequate mesh densities (with FEM), and, where confidence
or experience is lacking, a thorough testing of the robustness of the solution against changes in the geometry
or mesh density.

Now, the energy stored in the resonator's electromagnetic field is
$U = (1/2) \int \int \int \mu |\textbf{H}|^2 \textrm{d} \textrm{V} $,
where $\textbf{H}$ is the infinite-$Q$ solution generated by the PDE-solver,
and $\mu$ is the common permeability for the resonator's interior.
For resonators that are axisymmetric, the 3D volume integral
$\int \int \int \textrm{d} \textrm{V} $ over the resonator's interior reduces
to the 2D integral $\int \int (2 \pi x) \textrm{d} x \textrm{d} y$ over
its medial cross-section.
The surface current induced in the resonator's
enclosing perfect-electric wall is given by (see ref.~\cite{inan00}, page 205, for example)
$\textbf{J}_s = \textbf{H}_{\rm{t}} \equiv \textbf{n} \pmb{\times} \textbf{H}$,
where $\textbf{H}_{\rm{t}}$ is the tangential component of $\textbf{H}$ with respect
to the resonator's electric-wall boundary.

One now exploits (first-order) perturbation theory, and equates the current immediately stated above
with that which would be induced into the electric walls of a resonator, identical to
its loss-less antecedent, but for it having electric walls of \emph{finite} conductivity.
The equating of the two currents assumes that the lossy walls are nevertheless made
out of (or coated with) a sufficiently `good' conductor, such that
the change from loss-less to lossy does not significantly affect the
shapes of the resonator's modes. This will typically be the case for a cavity resonator
exciting low-order modes at microwave frequencies, provided its walls are made from
any standard (electrically good) metal, such as copper; again,
see references \cite{inan00} and/or \cite{bleaney76} for further explanation/quantification.

The time-averaged(-over-a-cycle) power lost by the resonator through resistive
heating in its imperfect electric walls is thereupon given by
$P_{\rm{loss}} = (1/2) \int \int  R_s |\textbf{n} \pmb{\times} \textbf{H}|^2 \textrm{d} \textrm{S} $,
where the 2D surface integral $\int \int \textrm{d} \textrm{S} $ over the resonator's presumed
axisymmetric enclosing boundary reduces to the 1D integral $\int (2 \pi x) \textrm{d} \textrm{l}$
around the periphery of its medial (x-y) cross-section;
$R_s = (\pi f \mu / \sigma)^{1/2}$ is the wall's surface resistivity,
where $\sigma$ is the wall's (bulk) electrically conductivity, and $f$ the mode's frequency.
The quality factor, defined as $2 \pi f \, U / P_{\rm{loss}}$, due to the wall's resistive losses
can thus be expressed as:
\begin{equation}
\label{eq:Q_wall_loss}
Q_{\rm{wall}}= \frac{2 \pi f \, \mu}{R_{\rm{s}}} \Lambda = (4 \pi f \, \mu \, \sigma)^{1/2} \Lambda,
\end{equation}
where $\Lambda$, which has the dimensions of a length, is defined as
\begin{multline}
\label{eq:length}
\Lambda = \frac{\int \int \int |\textbf{H}|^2 \textrm{d} \textrm{V} }
{\int \int |\textbf{n} \pmb{\times} \textbf{H}|^2 \textrm{d} \textrm{S} }\\
= \frac{\int \int [(H^x)^2 + (H^\phi)^2 + (H^y)^2] \, x \, \textrm{d} x \,\textrm{d} y}
{\int x \, [|H^\phi|^2 + |H^y dn_x - H^x n_y|^2] \rm{dl}}.
\end{multline}
Both integrals (numerator and denominator), hence  $Q_{\rm{wall}}$ itself, can be readily evaluated using
the PDE-solver's post-processing features; explicit PDE-solver-ready forms of each integrand are
provided in this paper's separate Appendix \cite{oxborrow06}.
In the using of equation \ref{eq:Q_wall_loss}, it should be pointed
out that, at liquid-helium temperatures, the bulk and surface resistances of metals can depend greatly
on the levels of (magnetic) impurities within them \cite{pobell92}, and the text-book $f^{-1/2}$ dependence
of surface resistance on frequency is often violated \cite{fletcher94}.

\section{Radiation Loss in Open Resonators\label{sec:RadLoss}}

\subsection{Open resonators: preliminary remarks\label{subsec:OpenGen}}
Many whispering-galley mode resonators (both microwave \cite{bourgeois04b} and optical
\cite{rokhsari05,srinivasan06}) of interest experimentally are \emph{not} shielded by
an enclosing metal wall: they are \emph{open}. In consequence, the otherwise highly localized WG mode
supported by the resonator spreads throughout free-space\footnote{As understood by Kippenberg \cite{kippenberg04},
this observation implies that the
support of equation \ref{eq:mode_volume}'s $\int \int \int_{\rm{b.-s.}} ... \textrm{d} \textrm{V} $ integral,
as it covers the WG mode's bright spot, must be somehow limited, spatially, or otherwise (asymptotically) rolled
off, lest the integral diverge as the so-called `quantization volume' associated
with the radiation extends to infinity.}, leading to the conveyance of energy
away from the mode's bright spot (where the electric- and magnetic-field amplitudes are greatest)
through \emph{radiation}.
Provided the equivalent closed resonator's enclosing electric (\emph{i.e.}~metal) wall is stationed sufficiently
far out in the WG mode's evanescent zone, the WG mode's form in its near-field will be the same
(to the degree of equivalence required here) in both the open-resonator and closed-resonator cases.
One can thus calculate the mode's near-field form through the method developed in section \ref{sec:Method},
or by some other method, as applied to the closed resonator; in particular, the electric and magnetic field
strengths, $\textbf{E}$ and $\textbf{H}$
or, equivalently, the vector potential $\textbf{A}$, just outside of the surface(s) of the resonator's
dielectric component(s) can be determined. Having done so, the WG mode's far-field form (\emph{i.e.}~its
`radiation pattern') in the case of the open resonator can be calculated by invoking the so-called
\emph{Field Equivalence Principle} \cite{schelkunoff36,balanis97}, where $\textbf{A}$ or (the tangential components of)
$\textbf{E}$ and $\textbf{H}$ over the said surface(s) are regarded, in Huygen's picture, as (secondary)
sources radiating into free-space. The calculation can be implemented through a standard
retarded-potential (Green function) approach \cite{ramo84,balanis97}, incorporating (if necessary) a
multipole expansion. The mode's radiative loss, hence $Q$, can be subsequently calculated from the radiation
pattern determined by integrating Poynting's vector over all solid angles. With due care,
the resulting estimate of the mode $Q$ will be highly accurate. But such a program of work
--for lack of novelty rather than utility-- shall \emph{not} be undertaken here.

\subsection{Estimators of radiation loss\label{subsec:RadiationEstimators}}
Instead, two different (but related) `trick' methods for \emph{estimating} the
radiative $Q$ of an open (dielectric) resonator are described here. As the first
method underestimates the $Q$, while the second overestimates it, the two in conjunction
can be used to bound the $Q$ from below and above. Moreover, both can be implemented
as straightforward `add-ons' to the 2D PDE-solver's computational environment,
as already configured for solving closed loss-less resonators (as per section \ref{sec:Method}).
It should be added that these two methods are not restricted to axisymmetric resonators \emph{per se}.

\subsubsection{Underestimator via (imperfect) retro-reflection\label{subsubsec:retro-refl}}
Consider an otherwise loss-less open resonator, supporting a spatially concentrated mode, \emph{i.e.}~one
with a bright spot, that radiates into free-space. As stated above, the tangential electric
and magnetic fields on any closed surface in the near-field surrounding this mode's
bright spot can, by the Field Equivalence Principle, be regarded as a (secondary) source of this radiation.
Now consider a closed, completely loss-less equivalent of the open resonator, formed by placing
a cavity around it, whose enclosing perfect-electric wall lies in the said localized mode's
radiation zone. It is noted here, for future reference, that this perfect electric wall
will force the tangential component of the electric field strength to vanish everywhere on it,
\emph{i.e.}~$\textbf{E} - \textbf{n} (\textbf{E} \cdot \textbf{n}) = 0$, where $\textbf{n}$ is the wall's
unit normal vector.
The above-mentioned secondary source generates an outward-going traveling wave wave which,
but for the cavity, would lead to radiation. Instead, with the cavity in place, a standing
(as opposed to traveling) wave arises.
Now, suppose that the shape of the cavity's electric wall, and its location with respect to the source,
is chosen to predominantly reflect the source's outward-going wave back to the source such that the resultant
inward-going wave interferes constructively with the outgoing wave over the whole of the source's surface.
In other words (1D analogy), on regarding the cavity as a short-circuit-terminated transmission line,
one attempts to adjust the length of the line such that its input (analogous to the source's surface)
is located at an antinode of the line's standing wave.
If such a retro-reflecting (+ phase-length adjusted) cavity can be devised then, in particular, the
measured/simulated tangential magnetic field, $\textbf{H}_{\rm{t}}$, just inside of the cavity's electric wall
will be exactly \emph{twice} that of the outward-going wave for the open resonator $\textbf{H}_{\rm{t}}^{\rm{rad.}}$
at the same location --but without the cavity's electric wall in place. In practice,
the source will not be located exactly at an antinode (over the whole of its surface) and thus
$\textbf{H}_{\rm{t}} > 2 \textbf{H}_{\rm{t}}^{\rm{rad.}}$.
The radiative loss for the open resonator can be evaluated by
integrating the cycle-averaged Poynting's vector corresponding
to the outward-going wave's inferred tangential magnetic field
over the electric wall's  surface;
\emph{i.e.} $P_{\rm{rad.}} = (1/2) \int \int  z_0 |\textbf{H}_{\rm{t}}^{\rm{rad.}}|^2 \textrm{d} \textrm{S} $,
where $z_0$ is the impedance of free space.
A bound on the open resonator's radiative $Q$-factor can thus be expressed as
\begin{equation}
Q_{\rm{rad.}} > (8 \pi f / c) \Lambda,
\label{eq:Q_rad_electric_wall}
\end{equation}
where $\Lambda$ is exactly that given by equation \ref{eq:length} but with the (loss-less) electric wall
now in the radiation zone. Provided the PDE solver is able to accurately calculate the (faint) electromagnetic
field on the rad.-zone cavity's enclosing boundary, it can again be readily determined.
It is further remarked here that the above --admittedly rather heuristic and one-dimensional-- argument, is strongly
reminiscent of Schelkunoff's induction theorem \cite{schelkunoff36,schelkunoff39}, which is itself a corollary of
the (already-mentioned) Field Equivalence Principle. Through analogy to this theorem,
the author conjectures that equation \ref{eq:Q_rad_electric_wall} holds equally well for fully
vectorial waves (as governed by Maxwell's equations) in 3D space as for scalar waves along 1D transmission
lines --as argued above.

\newpage
\subsubsection{Overestimator via (imperfect) outward-going free-space impedance match\label{subsec:Open}}
A complementary estimator to the one above can be constructed by replacing the above cavity's
electric wall with an `impedance-matched' one, where the tangential magnetic and tangential electric
fields at every point on the wall are constrained so as to correspond to those of an outward-going
plane \emph{traveling} wave, propagating in the direction of the wall's local normal and in
an outward-going sense. In other words (1D analogy), one attempts to confront the secondary
source's outward going wave with a matched surface that reflects nothing back.
For plane-wave radiation, this constraint can be expressed as
$z_0 \textbf{n} \pmb{\times} \textbf{H} = \textbf{E} - \textbf{n} (\textbf{E} \cdot \textbf{n})$,
where $\textbf{n}$ is the surface's inward-pointing normal. Note that one does not constrain
the direction (polarization) of $\textbf{E}$ or $\textbf{H}$ in the wall's local plane; one only demands that
the two fields be orthogonal and that their relative amplitudes be in the ratio of the impedance of
free space $z_0$. Upon differentiation w.r.t.~time and using Maxwell's displacement-current equation,
this relation can be re-expressed as
\begin{equation}
\label{eq:outward_rad_enf}
\nabla \pmb{\times} \textbf{H} - \textbf{n} [(\nabla \pmb{\times} \textbf{H}) \cdot  \textbf{n}]
- (1/c) \textbf{n} \pmb{\times} \frac{\partial \textbf{H}}{\partial t}= 0.
\end{equation}
For a given eigenmode, and generalizing somewhat, the constraint can be recast as
\begin{multline}
\label{eq:rad_match_mix}
\cos(\theta_{\rm{mix}}) \{\nabla \pmb{\times} \textbf{H} - \textbf{n} [(\nabla \pmb{\times} \textbf{H}) \cdot  \textbf{n}]\}\\
+ \sin(\theta_{\rm{mix}}) \, \textrm{i} \, \bar{c} f \, \textbf{n} \pmb{\times} \textbf{H} = 0,
\end{multline}
where $f$ is the mode's frequency (in Hz), $\bar{c} \equiv 2 \pi / c$ as before, and $\theta_{\rm{mix}}$ is
a `mixing angle'.
In the impedance-matched case (outward going plane wave in free space), $\theta_{\rm{mix}} = \pi/4$, and the above equation reduces to
\begin{equation}
\label{eq:rad_impedance_match}
\nabla \pmb{\times} \textbf{H} - \textbf{n} [(\nabla \pmb{\times} \textbf{H}) \cdot  \textbf{n}]
+ \textrm{i} \, \bar{c} f \, \textbf{n} \pmb{\times} \textbf{H} = 0 .
\end{equation}
More generally, the first and second terms on the left-hand side of equation
\ref{eq:rad_match_mix} can be viewed as representing electric- (cf.~equation \ref{eq:electricwallD})
and magnetic-wall (cf.~equation \ref{eq:magneticwallH}) boundary conditions, respectively,
where the latter corresponds to that used in subsubsection \ref{subsubsec:retro-refl} above.
The boundary condition can be
continuous adjusted between these two extrema by varying the mixing angle
$\theta_{\rm{mix}}$; for the sake of completeness, $\theta_{\rm{mix}} = -\pi/4$
corresponds to an inward-going, as opposed to an outward-going, impedance match.
Note that, unless $\theta_{\rm{mix}} = N \pi/2$ for integer $N$,
the square root of minus one in equation \ref{eq:rad_match_mix}
breaks the hermitian-ness of the matrix that the PDE-solver is required to eigensolve,
leading to solved modes with complex eigenfrequencies $f_{\rm{mode}}$.
As exploited by Srinivasan \emph{et al} \cite{srinivasan06},
the inferred quality factor for such a mode due to radiation through/into its bounding wall is given by
$Q_{\rm{inf.}} = \Re[f_{\rm{mode}}]/2\Im[f_{\rm{mode}}]$, where $\Re[...]$ and
$\Im[...]$ represent taking real and imaginary parts (of the complex eigenfrequency), respectively.

Note that the accuracy of the method will depend on the degree to which
the imposed surface impedance agrees with that of the true outward-going traveling
wave, as generated by the open resonator's (secondary) source, over the chosen
bounding surface.
If the source were an infinitessimal(finite) multi-pole,
then a surface in the form of a finite(infinite) sphere centered on the source,
with the constraint \ref{eq:rad_impedance_match} imposed on its surface,
would perfectly match to the source's radiation
(\emph{i.e.}~no traveling wave would get reflected back from it).
In general, however, for
a finite radiator, the chosen surface (necessarily of finite extent) will not lie
everywhere normal to the outward-going wave's Poynting's vector and back reflections
will result, leading to a smaller imaginary part in the simulated eigenmode's frequency,
thus causing $Q_{\rm{inf.}}$ to overestimate the true radiative $Q$. Thus, one may state
\begin{equation}
\label{eq:Q_rad_match}
Q_{\rm{rad.}} < \Re[f_{\rm{mode}}]/2\Im[f_{\rm{mode}}],
\end{equation}
approaching equality on perfect matching. Again, the author conjectures that, despite the
rather heuristic and one-dimensional argument stated above, inequality \ref{eq:Q_rad_match}
holds in general. Used together, equations \ref{eq:Q_rad_electric_wall} and \ref{eq:Q_rad_match}
provide a bounded estimator on the true radiative $Q$.

\emph{Comment:} As alluded to at the beginning of this section, the author recognizes that alternative (one
might argue rather more `empirical') approaches, based on surrounding (cladding) the otherwise
open resonator with sufficiently thick layers of impedance-matched absorber [\emph{i.e.}, with the absorber's
dielectric constant set equal to that of free space except for a small imaginary part (loss tangent)
causing the outward-going wave to be gently attenuated with little back reflection].
The `boundary-alteration' method described above has the advantage of not extending the
footprint of the PDE-solver's modeled region (thus \emph{not} requiring the mesh of finite
elements to be extended).

\section{Example applications\label{sec:ExampleApplications}}
The author has deployed the methodologies expounded in sections \ref{sec:Method} through \ref{sec:RadLoss} above
to model several different sorts of resonator. Where possible, he chose resonators with shapes and properties that
had already been published --so as to afford comparisons.  Each of the characteristics considered
in sections \ref{sec:Postprocessing} and \ref{sec:RadLoss} was evaluated for at least one such model resonator.
The COMSOL applications (as `.MPH' files) that the author constructed to simulate these resonators
can be obtained from him upon request.

\subsection{UWA `sloping-shoulders' cryogenic sapphire microwave resonator\label{subsec:SlopingShoulders}}
\begin{figure}[h]
\centering
\begin{tabular}{@{}r@{\quad}l}
a:$\:$\mbox{\includegraphics[width=0.42\columnwidth]{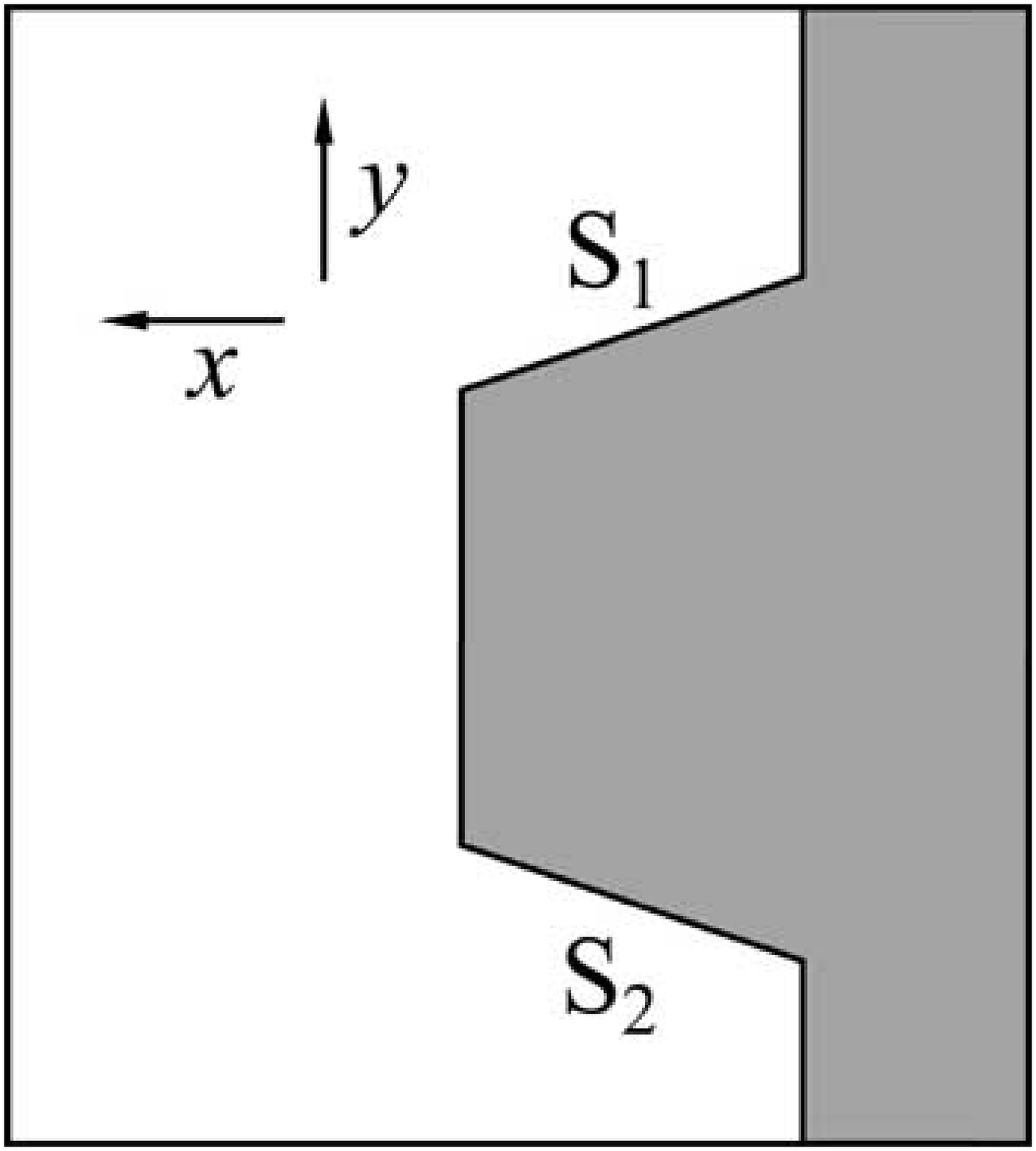}}&
b:$\:$\mbox{\includegraphics[width=0.42\columnwidth]{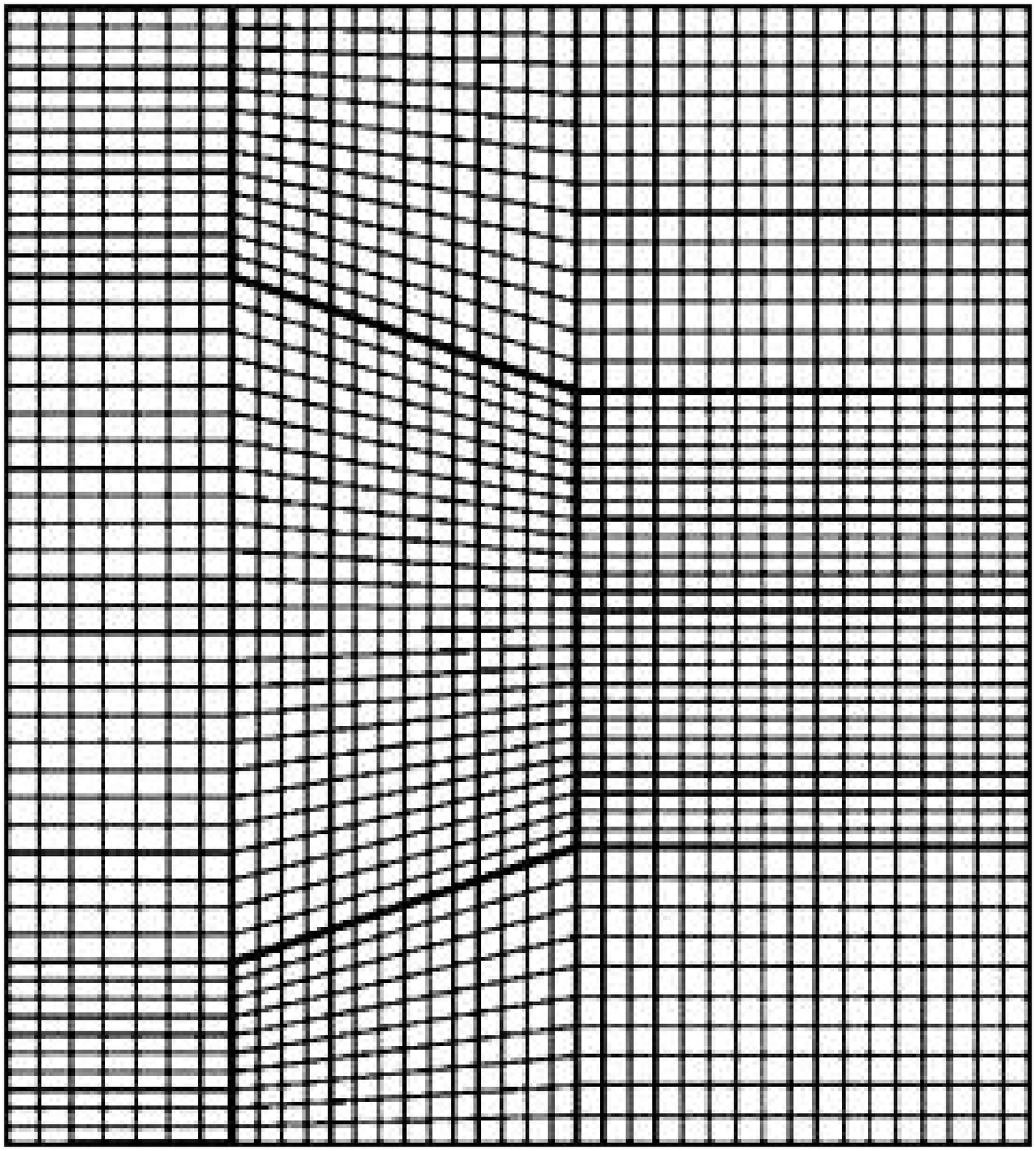}}\\
& \\
c:$\:$\mbox{\includegraphics[width=0.42\columnwidth]{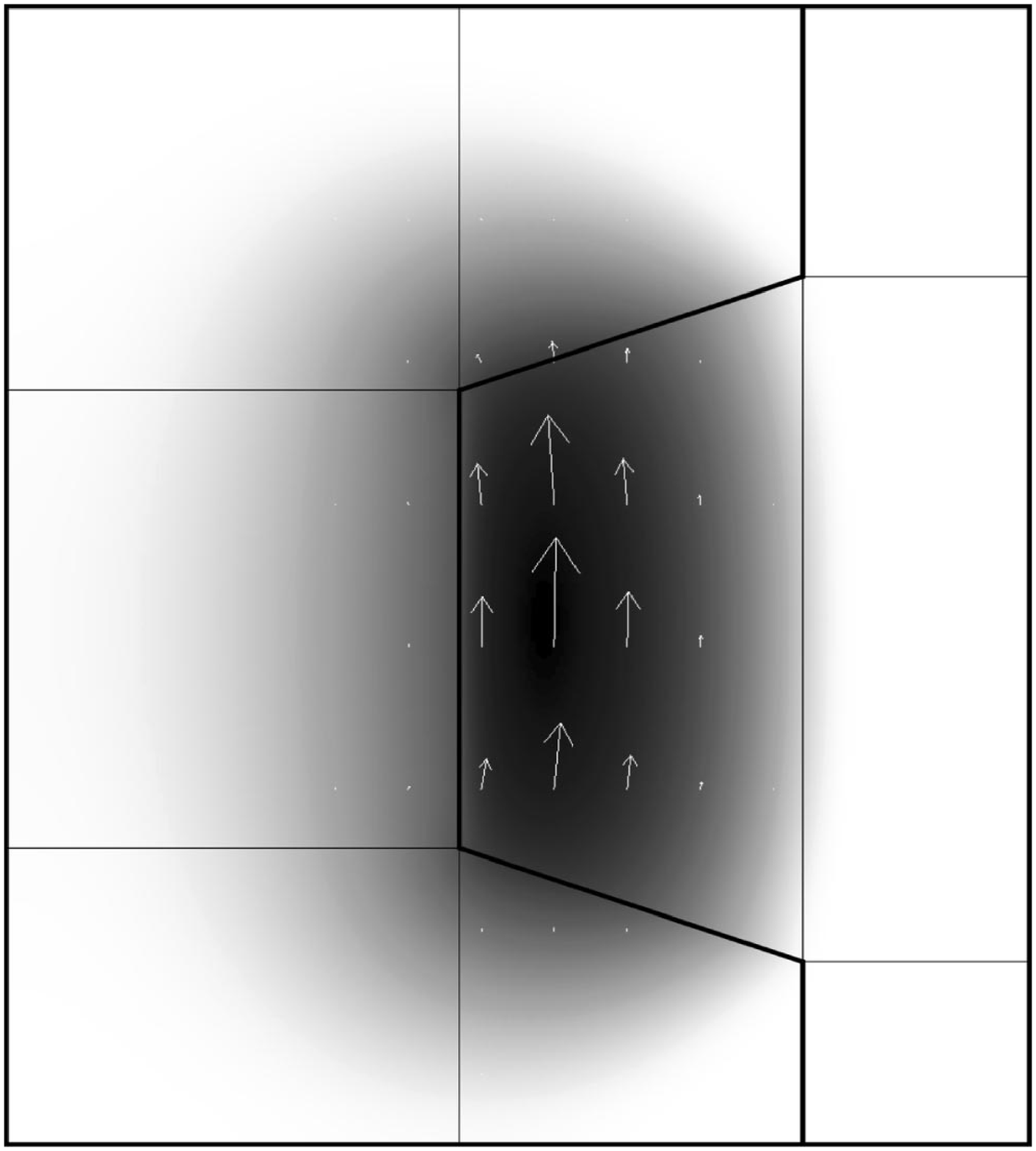}}&
d:$\:$\mbox{\includegraphics[width=0.42\columnwidth]{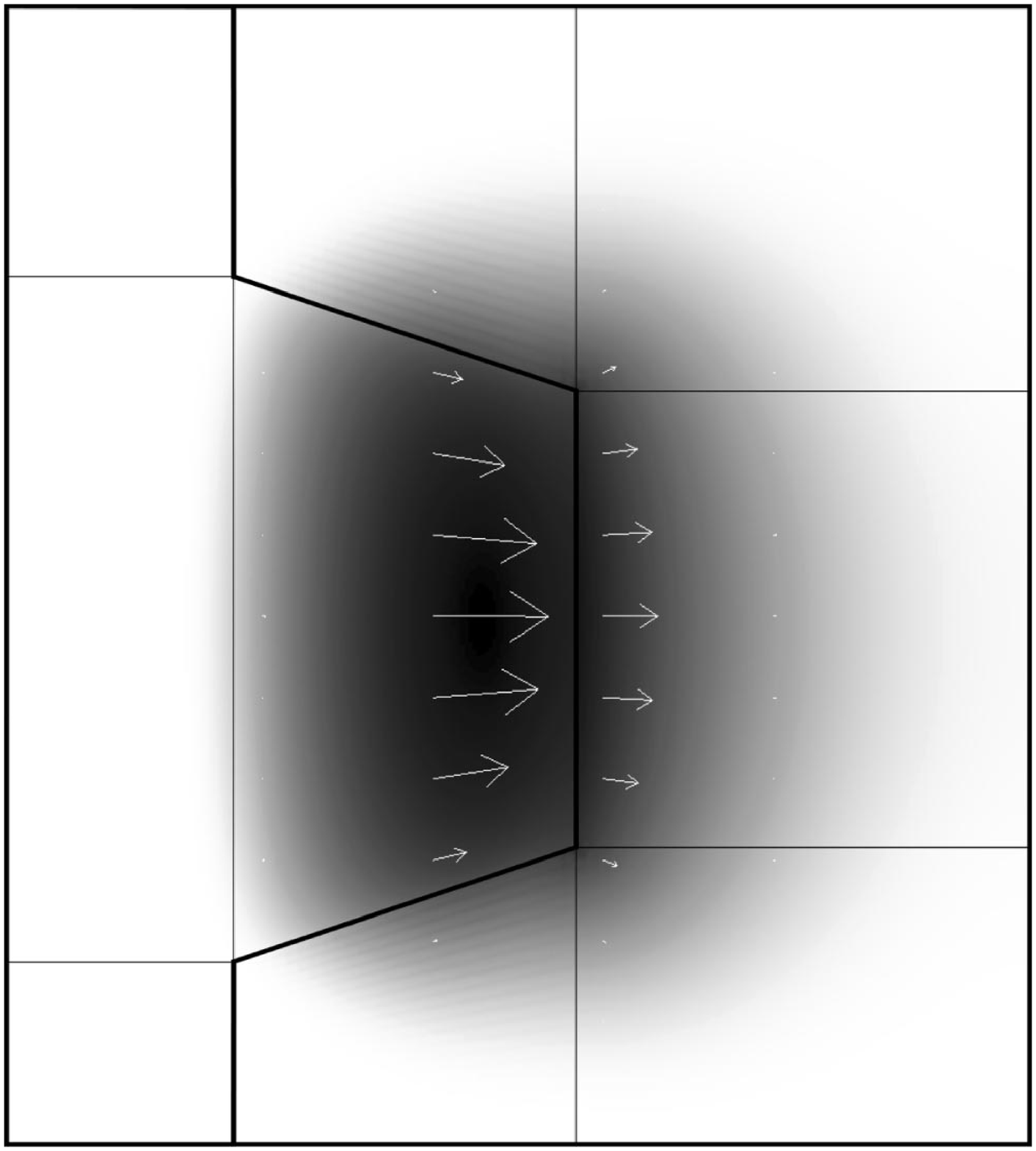}}\\
\end{tabular}
\caption{\footnotesize \it{UWA `sloping-shoulders' cryogenic sapphire resonator: (a) medial cross-section through it;
the grey(white) shading corresponds to sapphire(vacuum); \textrm{S}$_1$ and \textrm{S}$_2$ indicate the sapphire piece's two (upper and lower)
`shoulders'.
(b) the FEM-based PDE-solver's meshing of the
resonator's model structure; for clarity only every other meshing line is drawn [\emph{i.e.}~(b) displays
the `half-mesh']; (c) magnetic field intensity of the resonator's WGE$_{14,0,0}$ mode; (d) electric field intensity
of the same}; in both (c) and (d) faint white arrows indicate the direction of the magnetic and electric field,
respectively, in the medial plane; `intensity' here means the absolute value of the vectorial
$\textbf{H}$ (or, equivalently, $\textbf{B}$) field, displayed on a logarithmic grey scale --darker being more intense.}
\label{fig:UWAres}
\end{figure}

This resonator, as designed and assembled by workers at the University
of Western Australia (UWA) \cite{luiten95,wolf04},
comprises a piece of monocrystalline sapphire mounted within a cylindrical metal can.
The can's internal wall and the sapphire's outer surface exhibit rotational symmetry about a common axis.
Furthermore, the optical (or `c') axis of the sapphire crystal is, to good approximation,
oriented parallel to this geometric axis.
The resonator can thus be taken (and modeled) as being electromagnetically axisymmetric.
The sapphire piece's medial cross-section (one half thereof) is shown in Fig.~\ref{fig:UWAres}(a). What makes the
resonator awkward to simulate accurately via the semi-analytic MM-SV method \cite{wolf04} is its sloping shoulders
(S$_1$ and S$_2$ \emph{ibid.}), whose surface normals are neither purely axial nor purely radial.

The resonator's form, as the author encoded it into the PDE-solver, is based on figure 3 of
ref.~\cite{wolf04}\footnote{It is commented parenthetically here that the shape of the sapphire piece in
figure 3 of ref.~\cite{wolf04} is not consistent with the dimensions stated in the same: its outer axial
sidewall is too long and the slope of its shoulders too small with respect to the stated axial dimensions.}.
For the simulation presented here, the author took the piece's outer diameter, the length of its outer
axial sidewall, the axial extent of each sloping shoulder, and the radius of each of its two spindles
to be, at liquid-helium temperature (\emph{i.e.}~the dimensions here stated include cryogenic
shrinkages --see section \ref{sec:PermDet}) 49.97, 19.986, 4.996, and 19.988~mm, respectively.
The sapphire crystal's cryogenic permittivities were taken to be $\{\epsilon_\perp ,
\epsilon_\parallel \} = \{9.2725, 11.3486\}$, as stated in ref.~\cite{krupka99a}.
Note that, though coaxial, the sapphire piece and the can do not exactly share a common transverse
(`equatorial') mirror plane, thus precluding any speeding up of the simulation through the
placement (in the model) of a magnetic or an electric wall on the equatorial plane,
thereupon halving the 2D region to be analyzed\footnote{These two boundary conditions
would lead to symmetric (N) and antisymmetric (S) solutions, respectively --see ref.~\cite{tobar01}.}.

Fig.~\ref{fig:UWAres}(b) displays the FEM-based PDE-solver's meshing of the resonator structure.
Here, the resonator's interior dielectric domains were regularly decomposed into quadrilaterals (as
opposed to triangles), with no quadrilaterals straddling interfaces between different materials. The mesh comprised,
in COMSOL's vernacular\footnote{The size/complexity of a finite-element mesh is quantified, within COMSOL Multiphysics,
by (i) the number of elements that go to compose its so-called `base mesh' and (ii) its total number of degrees
of freedom (`DOF') --as associated with its so-called `extended mesh'. Consult the package's documentation for
further clarifications.}, 7296 base-mesh elements and 88587 degrees of freedom (`DOF').
It typically took around 75~seconds, to obtain the resonator's lowest (in frequency) 16 modal solutions,
for a single, given azimuthal mode order $M$, at [with respect to Fig.~\ref{fig:UWAres}(b)] full mesh
density, on a standard, 2004-vintage personal computer (2.4~GHz, Intel Xeo CPU), without altering the PDE-solver's
default eigensolver settings. With the azimuthal mode order set at $M = 14$, the model resonator's WGE$_{\rm{14,0,0}}$ mode was
found to lie at 11.925 GHz, to be compared with 11.932~GHz found experimentally \cite{wolf04}.

\noindent\emph{Wall loss:}
This mode's characteristic length $\Lambda$ was determined to be $2.6 \times 10^4$.
Based on ref.~\cite{fletcher94}, one estimates the surface resistance of copper at liquid-helium temperature
to be $7 \times 10^{-3} \; \Omega$ per square at 11.9 GHz, leading to a
wall-loss $Q$ of  $3.5 \times 10^{11}$ for the  WGE$_{\rm{14,0,0}}$ mode. As this is at least an order of magnitude
greater than what is observed experimentally, one concludes that wall losses do not substantially limit the UWA resonator's
experimental $Q$.

\noindent \emph{Filling factor:} Using equation \ref{eq:filling_factor}, the electric filling factors for the WGE$_{14,0,0}$ mode have been
evaluated. The results, presented in TABLE~\ref{tab:UWAfillfac}, are to be compared with those stated in
Table~IV of ref.~\cite{wolf04}: they are in good agreement with those \emph{loc. cit.} (labeled `FE'), which
were obtained via a wholly independent finite-element analysis.
\begin{table}[t]
\centering
\caption{Electric filling factors for the WGE$_{\rm{14,0,0}}$ mode of the UWA resonator}
\begin{tabular}{@{}r|lll}
$F_{\rm{diel.}}^{\rm{dir.}}$        & radial        & azimuthal                 & axial \\
\hline
sapphire                            & 0.80922       & 0.16494                   & $7.016 \times 10^{-3}$ \\
vacuum                              & 0.01061       & $8.0533 \times 10^{-3}$   & $1.6543 \times 10^{-4}$ \\
\end{tabular}
\label{tab:UWAfillfac}
\end{table}

\subsection{Toroidal silica optical resonator [Caltech]\label{subsec:SilicaToroidal}}
\begin{figure}[h]
\centering
\begin{tabular}{@{}l}
a:$\:$ \mbox{\includegraphics[width=0.75\columnwidth]{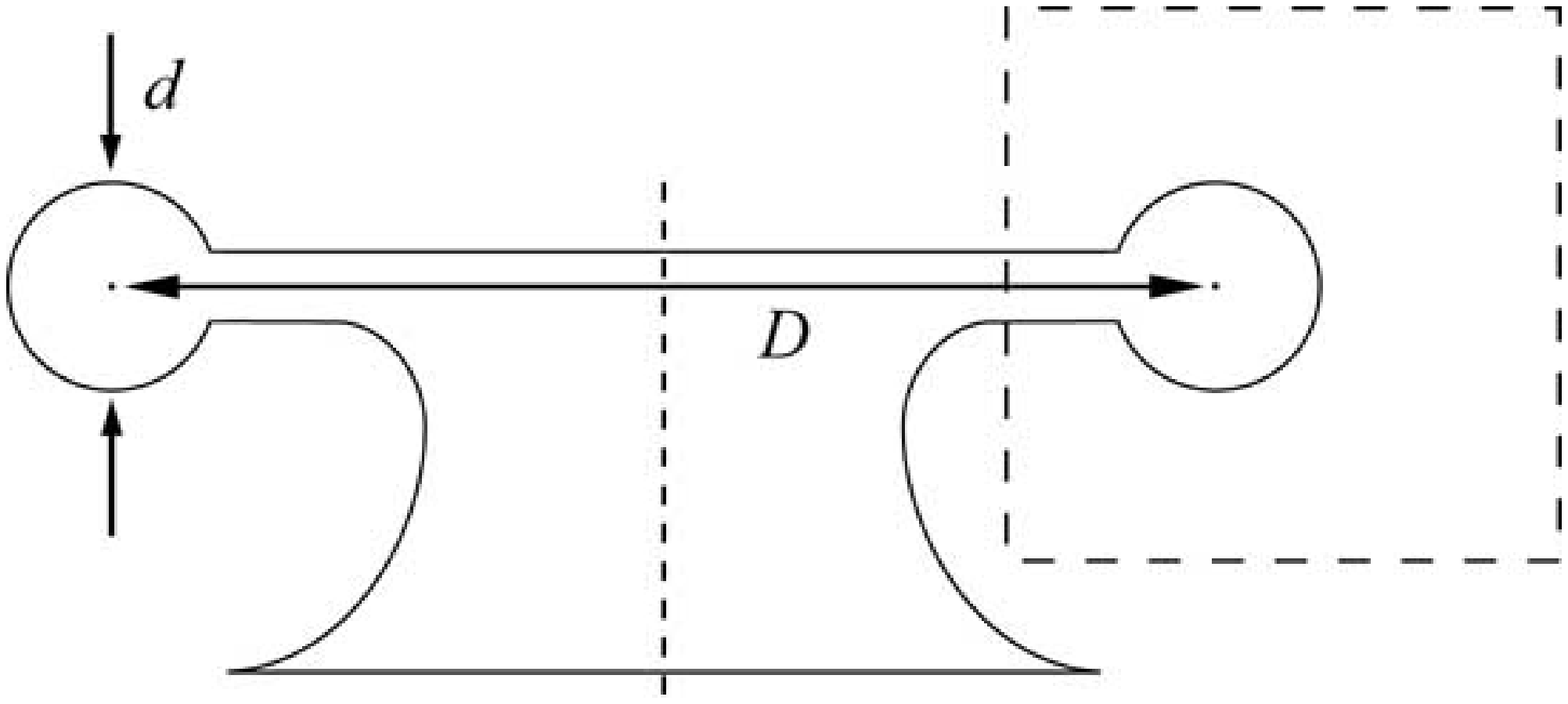}}\\
\\
b:$\:$ \mbox{\includegraphics[width=0.55\columnwidth]{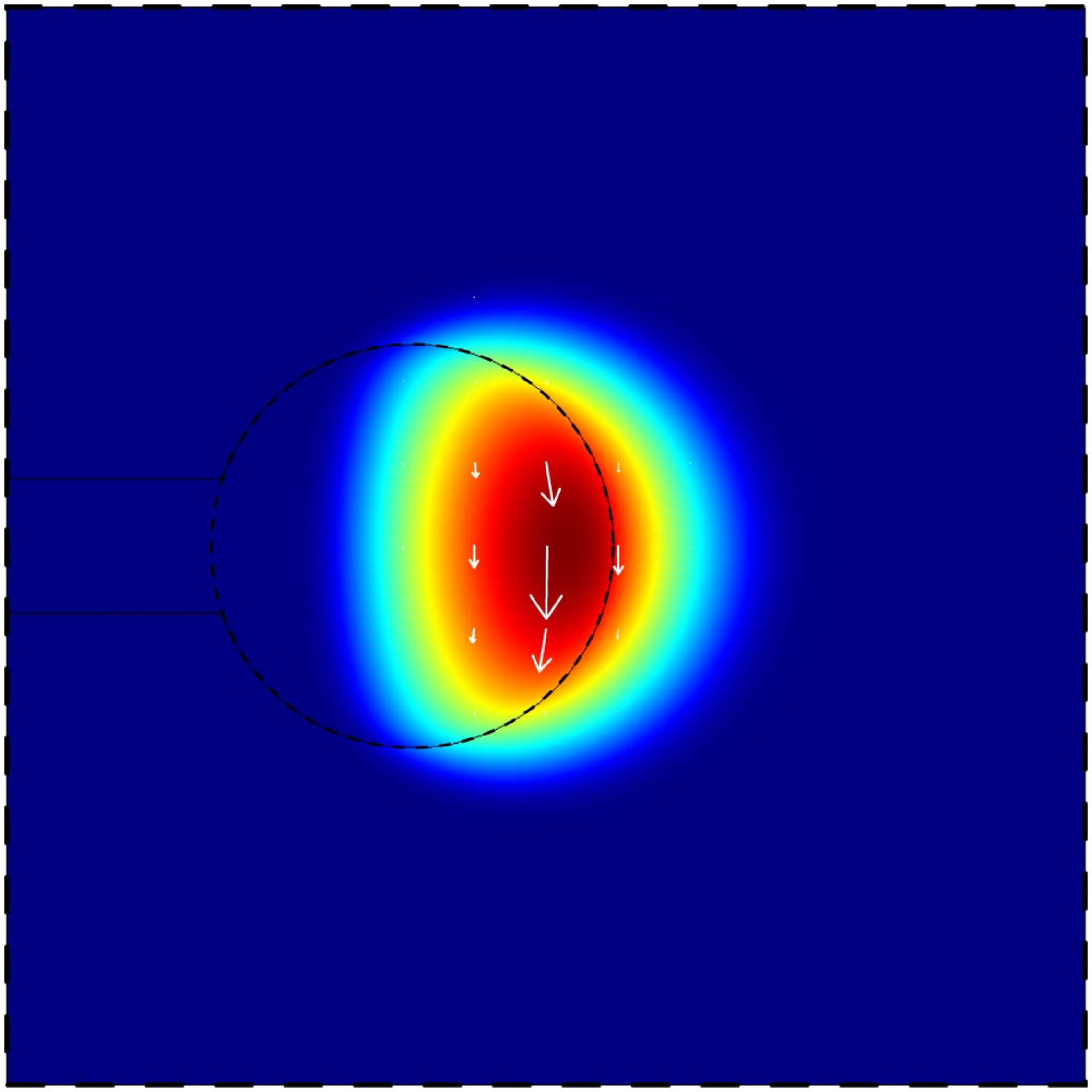}}\\
\end{tabular}
\caption{(a) Geometry (medial cross-section) and dimensions of a model toroidal silica microcavity resonator
--after ref. \cite{spillane05}; the torus' principal diameter $D = 16$ $\mu$m and its
minor diameter $d = 3$ $\mu$m; the central vertical dashed line indicates the resonator's axis of continuous
rotational symmetry.
(b) False-color surface plot of the (logarithmic) electric-field intensity $|\textbf{E}|^2$ within the dashed
box appearing in (a) for this resonator's TE$_{p=1,m=93}$ whispering-gallery mode. White arrows indicate
the electric field $\bf{E}$'s magnitude and direction in the medial plane.}
\label{fig:ToroidalSilica}
\end{figure}

The resonator modeled here, based on ref.~\cite{spillane05}, comprises a silica
toroid, where this toroid is supported through an integral interior `web', such that the
toroid is otherwise suspended in free space above the resonator's substrate.
This arrangement is shown in Fig.~\ref{fig:ToroidalSilica}(a).
The toroid's principal and minor diameters
are set at $\{D,d\}$ = $\{16,3\}$ $\mu$m, respectively. The silica dielectric is presumed
to be wholly isotropic (\emph{i.e.}, no significant residual stress) with a relative permittivity
of $\epsilon_{\rm{sil.}} = 2.090$, corresponding to a refractive index of
$n_{\rm{sil.}} = \sqrt{\epsilon_{\rm{sil.}}} = 1.4457$ at the resonator's
operating wavelength (around 852~nm) and temperature.
The FEM model's mesh covered an 8-by-8 $\mu$m square [shown in dashed outlined
on the right of Fig.~\ref{fig:ToroidalSilica}(a)] in the medial half-plane
containing the silica toroid's circular cross-section.
A pseudo-random triangular mesh was generated (automatically)
with an enhanced meshing density over the silica circle and
its immediately surrounding free-space; in total,
the mesh comprised 5990 (base-mesh) elements, with DOF $= 36279$.
Temporarily adopting Spillane \emph{et al}'s terminology, the resonator's fundamental TE-polarized
93rd-azimuthal-mode-order mode (where by `TE' it is here meant that the polarization of the
mode's electric field is predominantly aligned with the toroid's principal axis --\emph{not} transverse to it)
was found to have a frequency of $3.532667 \times 10^{14}$~Hz, corresponding to a
free-space wavelength of $\lambda$ = 848.629 nm (thus close to 852 nm).
Using this paper's equation \ref{eq:mode_volume}, this mode's volume was evaluated
to be 34.587 $\mu$m$^3$; if, instead, the definition stated in equation 5 of
ref.~\cite{spillane05} is used, the volume becomes
72.288 $\mu$m$^3$ --\emph{i.e.}~a factor of $n_{\rm{sil.}}^2$ greater.
These two values straddle (neither agree with) the volume
of approx.~55~$\mu$m$^3$, for the same dimensions of silica toroid
and the same (TE) mode-polarization, as inferred by eye-and-ruler from
figure 4 of ref.~\cite{spillane05}. The author cannot explain the discrepancy.

It is pointed out here that the white arrows in Fig.~\ref{fig:ToroidalSilica} (at least those not anchored on
the equatorial plane) are slightly but noticeably oriented away from vertical, indicating that the orientation
of the mode's (vectorial) electric field is not perfectly axial --as per the transverse approximation taken
in references \cite{wolf04,kippenberg04}.
In other words, the arrows' lack of verticality reveals the inexact- or \emph{quasi}-ness
of the mode's transverse-electric nature, despite the mode's relatively high azimuthal
mode order ($l \equiv M = 93$).
\subsection{Conical microdisk optical resonator [Caltech]\label{subsec:GaAlAsMicrodisk}}
The mode volume can be reduced by going to smaller resonators, which, unless
the optical wavelength can be commensurately reduced, implies \emph{lower} azimuthal mode order.
\begin{figure}[h]
\centering
\begin{tabular}{@{}l}
a:$\:$ \mbox{\includegraphics[width=0.75\columnwidth]{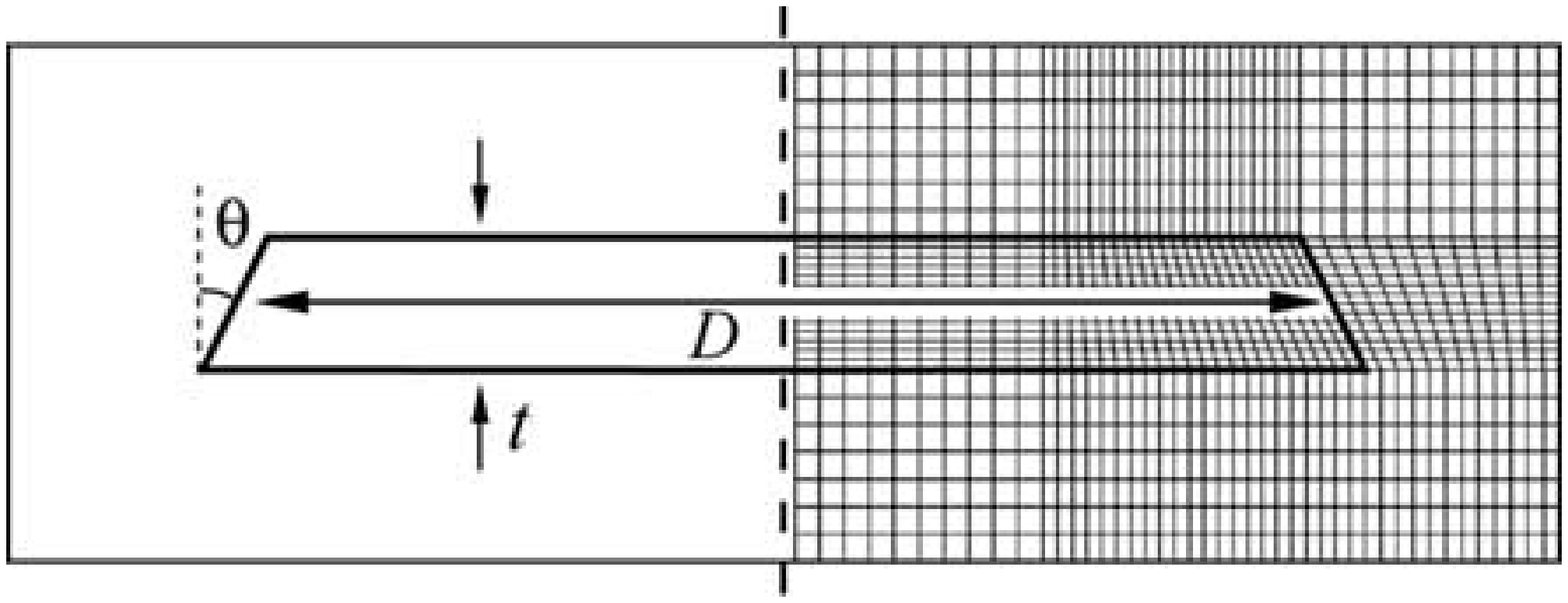}}\\
\\
b:$\:$ \mbox{\includegraphics[width=0.65\columnwidth]{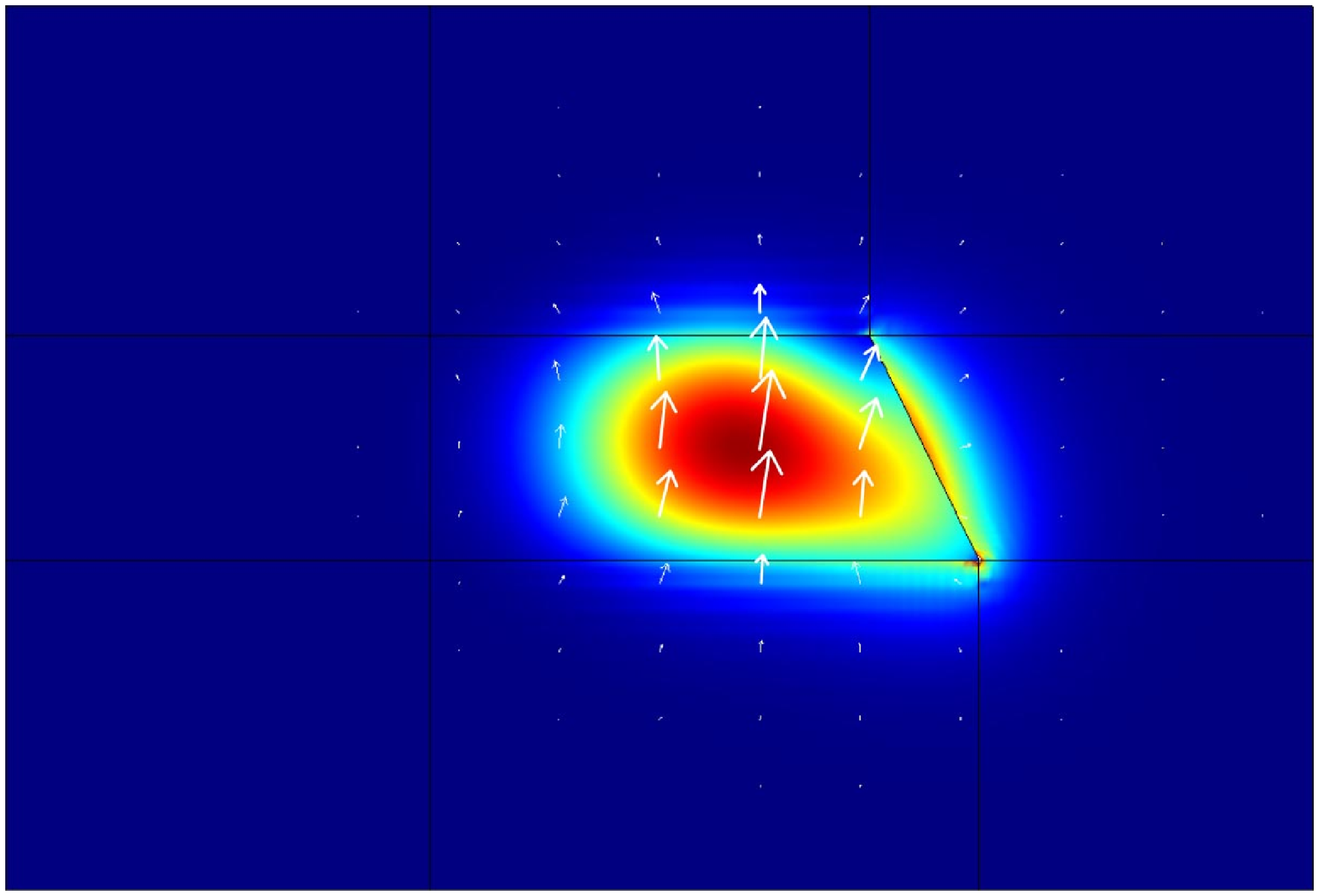}}\\
\end{tabular}
\caption{(a) Geometry (medial cross-section) and (half-)meshing of model GaAlAs microdisk resonator
--after ref. \cite{srinivasan06}; the disk's median diameter is $D$ = 2.12 $\mu$m and its
thickness (axial height) $t$ = 255 nm; its conical external sidewall subtends an angle
$\theta$ = 26$^\circ$ to the disk's (vertical) axis; for clarity, only every other line of the true
(full) mesh is drawn. The modeled domain in the medial half-plane is a rectangular stretching from
0.02 to 1.5 $\mu$m in the radial direction and from -5 to +5 $\mu$m in the axial direction.
(b) False-color surface plot of the (logarithmic) electric-field intensity
$|\textbf{E}|^2$ for the resonator's TE$_{p=1,m=11}$ mode at $\lambda$ = 1263.6 nm. Again,
white arrows indicate the electric field's magnitude and direction in the medial plane.}
\label{fig:GaAlAsMicrodisk}
\end{figure}
The model `microdisk' resonator analyzed here, as depicted in Fig.~\ref{fig:GaAlAsMicrodisk}(a)
is the author's attempt at duplicating the structure
defined in figure 1(a) of Srinivasan \emph{et al}~\cite{srinivasan06}; as in their model,
the disk's constituent dielectric (in reality, layers of GaAs and GaAlAs) is
approximated as a spatially uniform, isotropic dielectric, with a refractive index equal to $n$ = 3.36.
The FEM-modeled domain in the medial half-plane was divided into 4928 quadrilateral
base-mesh elements, with DOF~$=60003$. Adopting the same authors' notation,
the resonator's TE$_{p=1,m=11}$ whispering-gallery mode,
as shown in Fig.~\ref{fig:GaAlAsMicrodisk}(b), was found
at $2.372517 \times 10^{14}$ Hz, equating to $\lambda$ = 1263.6~nm;
for comparison, Srinivasan \emph{et al} found an equivalent mode at 1265.41~nm
[as depicted in their figure 1(b)]. The white arrows' lack of verticality
in this article's Fig.~\ref{fig:GaAlAsMicrodisk}(b) implies that the orientation
(\emph{i.e.}~polarization) of the magnetic field associated with the true,
quasi-TE$_{p=1,m=11}$ mode deviates significantly from axial
(as would be the case within a transverse approximation).

\noindent \emph{Mode volume:}
Using this paper's equation \ref{eq:mode_volume}, but with the mode excited as a
standing-wave (doubling the numerator while quadrupling the denominator),
the mode volume is determined to be $0.1484 \times \mu$m$^3 \simeq 2.79 (\lambda/n)^3$,
still in good agreement with Srinivasan \emph{et al}'s $\sim \! 2.8 (\lambda /n)^3$.

\begin{figure}[h]
\centering
\begin{tabular}{@{}l@{\quad}l@{\quad}l}
a:$\:$ \mbox{\includegraphics[width=0.25\columnwidth]{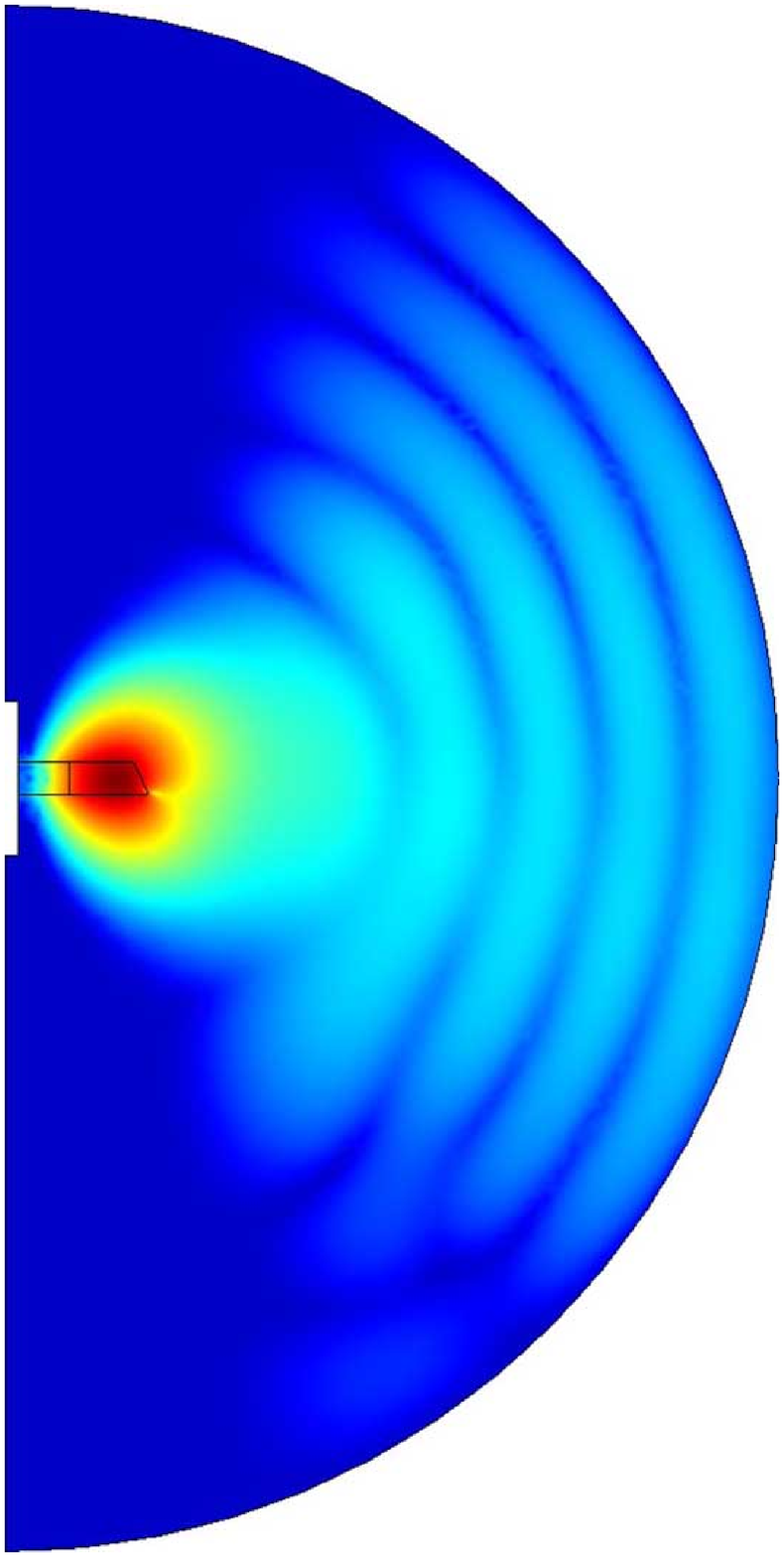}}
& b:$\:$ \mbox{\includegraphics[width=0.25\columnwidth]{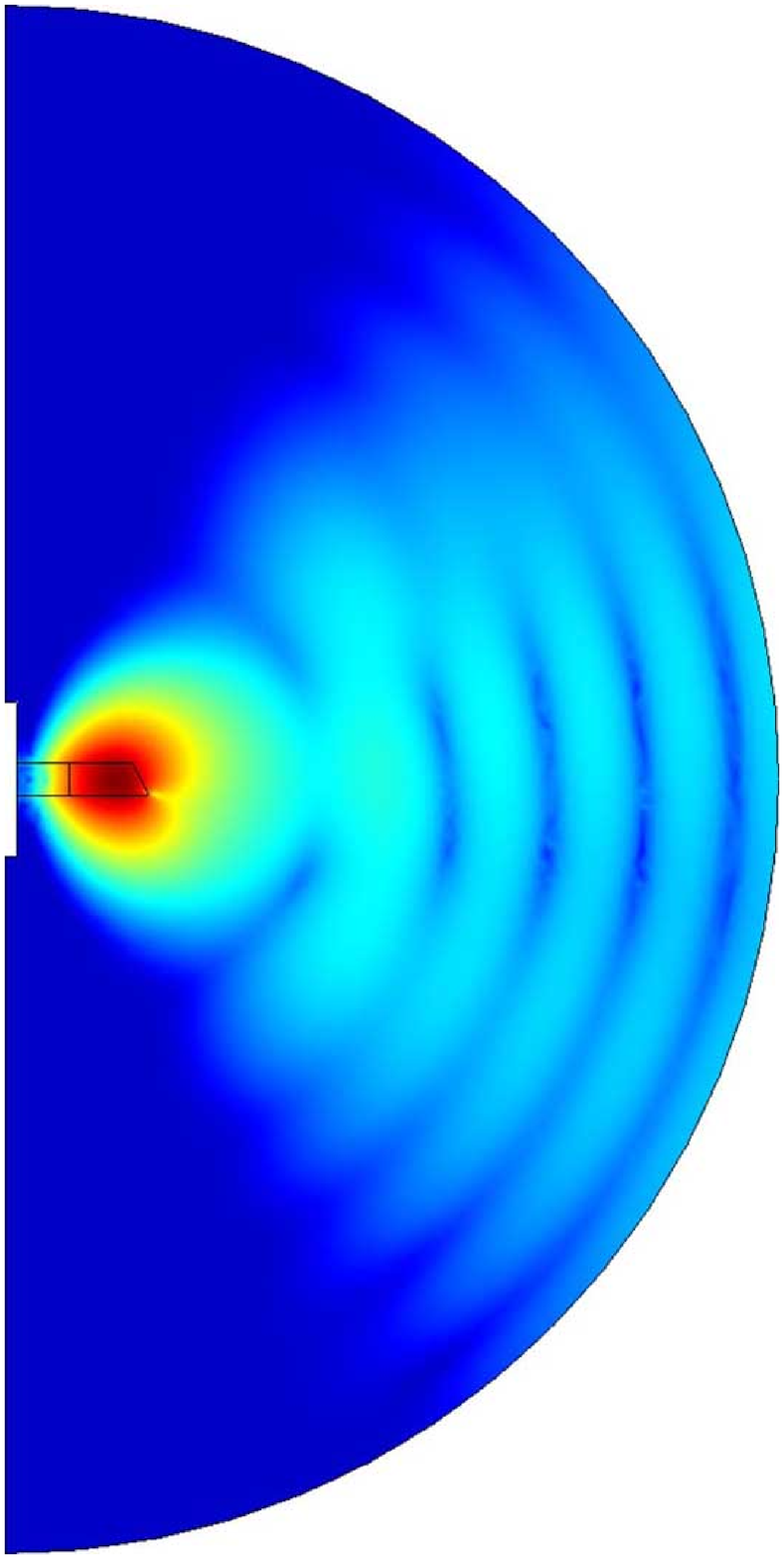}}
& c:$\:$ \mbox{\includegraphics[width=0.25\columnwidth]{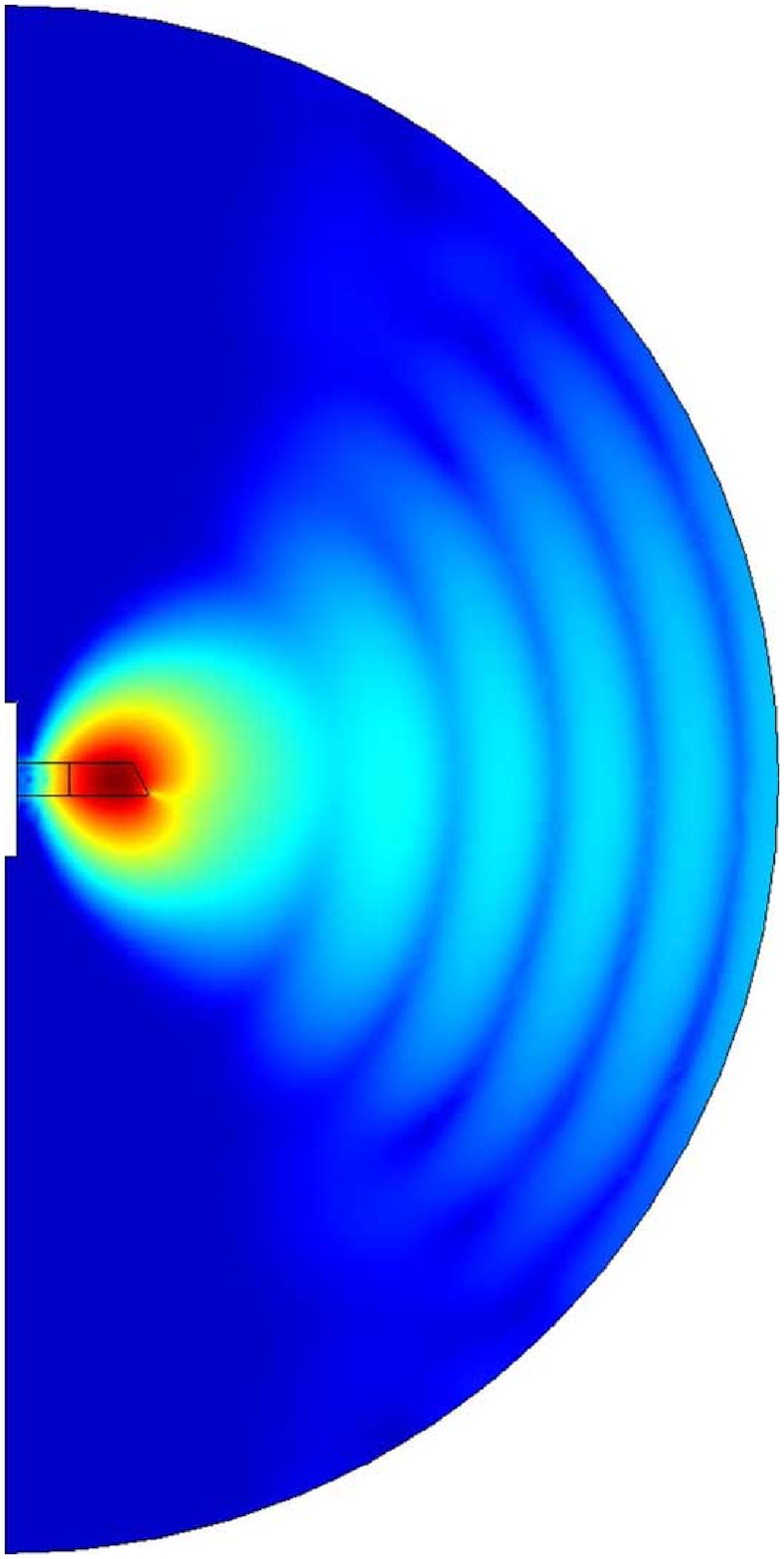}}\\
\end{tabular}
\caption{Radiation associated with the same [TE$_{p=1,m=11}$, $\lambda$ = 1263.6 nm] whispering-gallery mode
as presented in Fig.~\ref{fig:GaAlAsMicrodisk};
here, false-color maps of the squared magnitude of the mode's magnetic field strength are plotted out
to the modeled domain's near-spherical outer boundary, corresponding in the medial half-plane to
a semi-circle 12 $\mu$m in diameter, whose center lies at a radial coordinate of 0.01 $\mu$m on the
microdisk's mid-plane.
[In reality, the microdisk's substrate would occupy a considerable part of the meshed half-disk's lower quadrant,
but the model here assumes that, with the exception of the microdisk itself (a dielectric), both quadrants are
filled with free space --into which the the whispering gallery mode radiates.] All three maps use the same
absolute false-color scale.
(a) standing-wave (equal outward- and inward-going) radiation with the outer semicircular boundary set as a
magnetic wall; (b) the same but now with the boundary set as an electric-wall; (c)
somewhat traveling (more outward- than inward-going) radiation with the boundary's impedance
set to that of an outward-going plane-wave in free space (and with the normal magnetic field
constrained to vanish). That (c)'s radiation field is somewhat dimmer than (b)'s is consistent
with the different  estimates of the resonator's radiative $Q$ corresponding to (a)-(c) [see text].}
\label{fig:RadiationLoss}
\end{figure}

\noindent \emph{Radiation loss:}
The TE$_{p=1,m=11}$ mode's radiation loss was estimated by implementing both the upper- and
lower-bounding estimators described in subsection \ref{subsec:RadiationEstimators}.
Here, the microdisk and its mode were modeled within a near-spherical
volume (equating to a half-disk in the medial half-plane, with a semicircle
for its outer perimeter), on whose outer boundary different electromagnetic
conditions were imposed --see Fig.~\ref{fig:RadiationLoss}.
With an electric-wall condition (\emph{i.e.} equations \ref{eq:electricwallH}
and \ref{eq:electricwallD} or, equivalently, \ref{eq:electricwallHcylcomp}
and \{\ref{eq:electricwallEcylcomp1}, \ref{eq:electricwallEcylcomp2isotrop}\})
imposed on the volume's whole boundary [as per Fig.~\ref{fig:RadiationLoss}(b)],
the right-hand of equation \ref{eq:Q_rad_electric_wall} was evaluated.
And, with the $\textbf{E} \pmb{\times} \textbf{n} = 0$ condition (\emph{viz.} equation \ref{eq:electricwallD})
on its outer semi-circle replaced by the outward-going-plane-wave(-in-free-space)
impedance-matching condition (\emph{viz.}~equation \ref{eq:rad_match_mix}), while the
$\textbf{H} \cdot \textbf{n} = 0$ condition (equation \ref{eq:electricwallH}) is maintained,
the right-hand side of equation \ref{eq:Q_rad_match} was evaluated for the radiation pattern displayed
in Fig.~\ref{fig:RadiationLoss}(c).
For a pseudo-random triangulation mesh comprising 4104 elements, with a DOF of 24927,
the PDE solver took, on the author's
office computer, 6.55 and 13.05 seconds, corresponding to
Figs.~\ref{fig:RadiationLoss}(b) and (c), respectively\footnote{The complex arithmetic
associated with the impedance-matching boundary condition meant that the PDE solver's
eigen-solution took approximately twice as long to run with this condition imposed --as compared to the electric-
(or magnetic-) wall boundary conditions that do not involve complex arithmetic.}, to calculate 10 eigenmodes around
$2.373 \times 10^{14}$ Hz, of which the TE$_{p=1,m=11}$ mode was one.
Together, the resultant estimate on the TE$_{p=1,m=11}$ mode's radiative-loss quality factor is
$(1.31 < Q_{\rm{rad.}} < 3.82) \times 10^7$,
to be compared with the estimate of
$9.8 \times 10^6$ (at 1265 nm) reported in table 1 of ref.~\cite{srinivasan06}\footnote{The author
chose the diameter of the outer semicircular boundary in
Figs.~\ref{fig:RadiationLoss}(a)-(c) arbitrarily to be 12.0 $\mu$m \emph{in advance
of} knowing what upper and lower bounds on $Q_{\rm{rad.}}$ such a choice would give;
he did not subsequently adjust the diameter and/or shape of this boundary
to bring the bounds any closer together.}.
The standing-wave radiation field in Fig.~\ref{fig:RadiationLoss}(b) could have been made dimmer (thus
increasing $\Lambda$, hence the inferred $Q$) by adjusting (`tuning') the meshed half-disk's diameter
--so as to put the microdisk/near-field TE$_{p=1,m=11}$ mode, viewed as a secondary source of radiation,
closer to an antinode of the cavity's standing-wave field.
Also, the simulations associated with Fig.~\ref{fig:RadiationLoss} could certainly
have run (with tolerable execution times) on a denser finite-element mesh.

\subsection{3rd-order Bragg-cavity alumina:air microwave resonator\label{subsec:BraggCav}}
Commercial FEM-based PDE-solvers (\emph{viz.}~the COMSOL/FEMLAB package used by the author for this article)
permit the simulation of arbitrarily complex structures and, moreover, provide efficient languages and tools
for representing and constructing (and modifying) them.
\begin{figure}[h]
\centering
\begin{tabular}{@{}l}
a:$\:$ \mbox{\includegraphics[width=0.5\columnwidth]{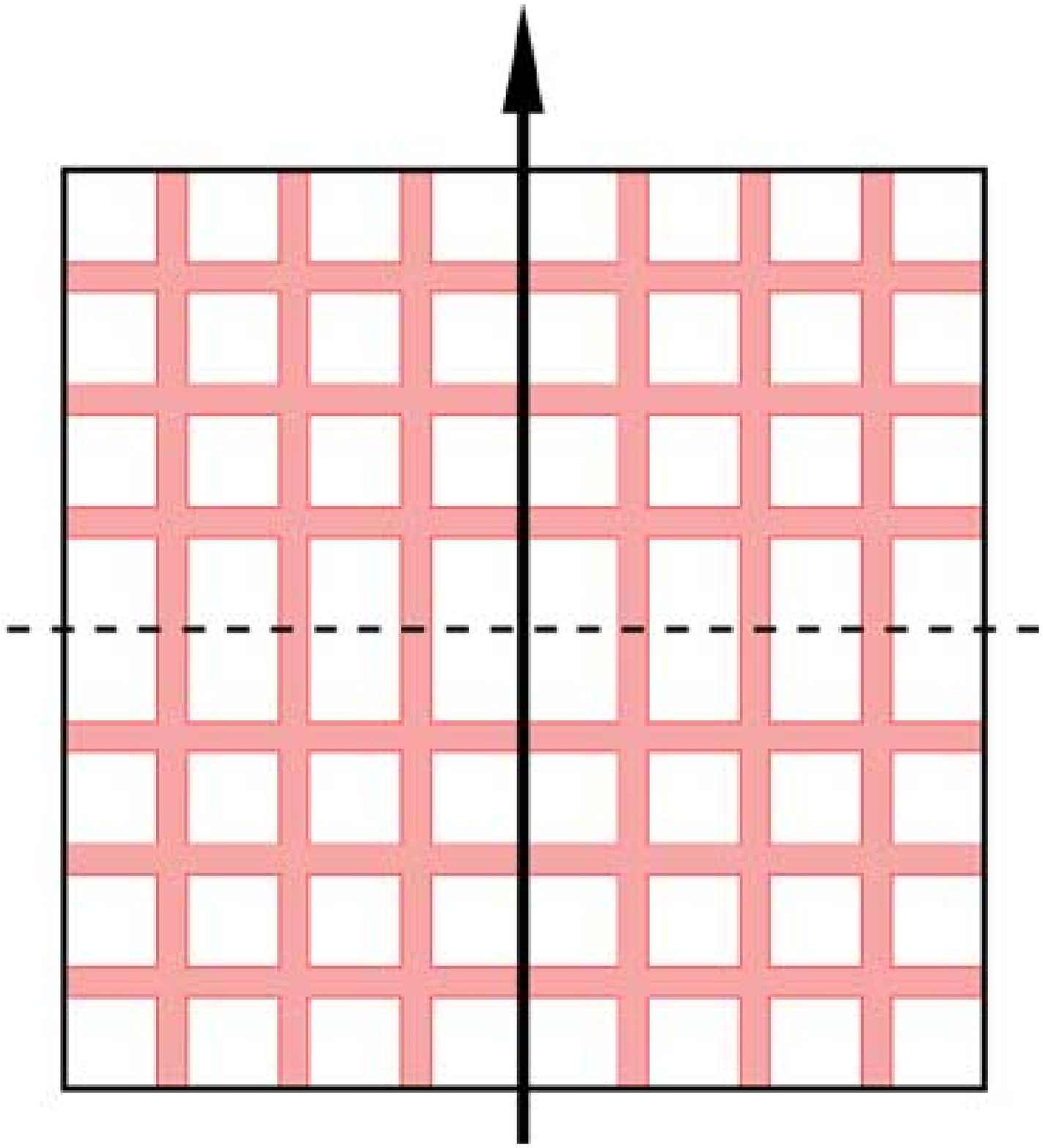}}\\
\\
b:$\:$ \mbox{\includegraphics[width=0.45\columnwidth]{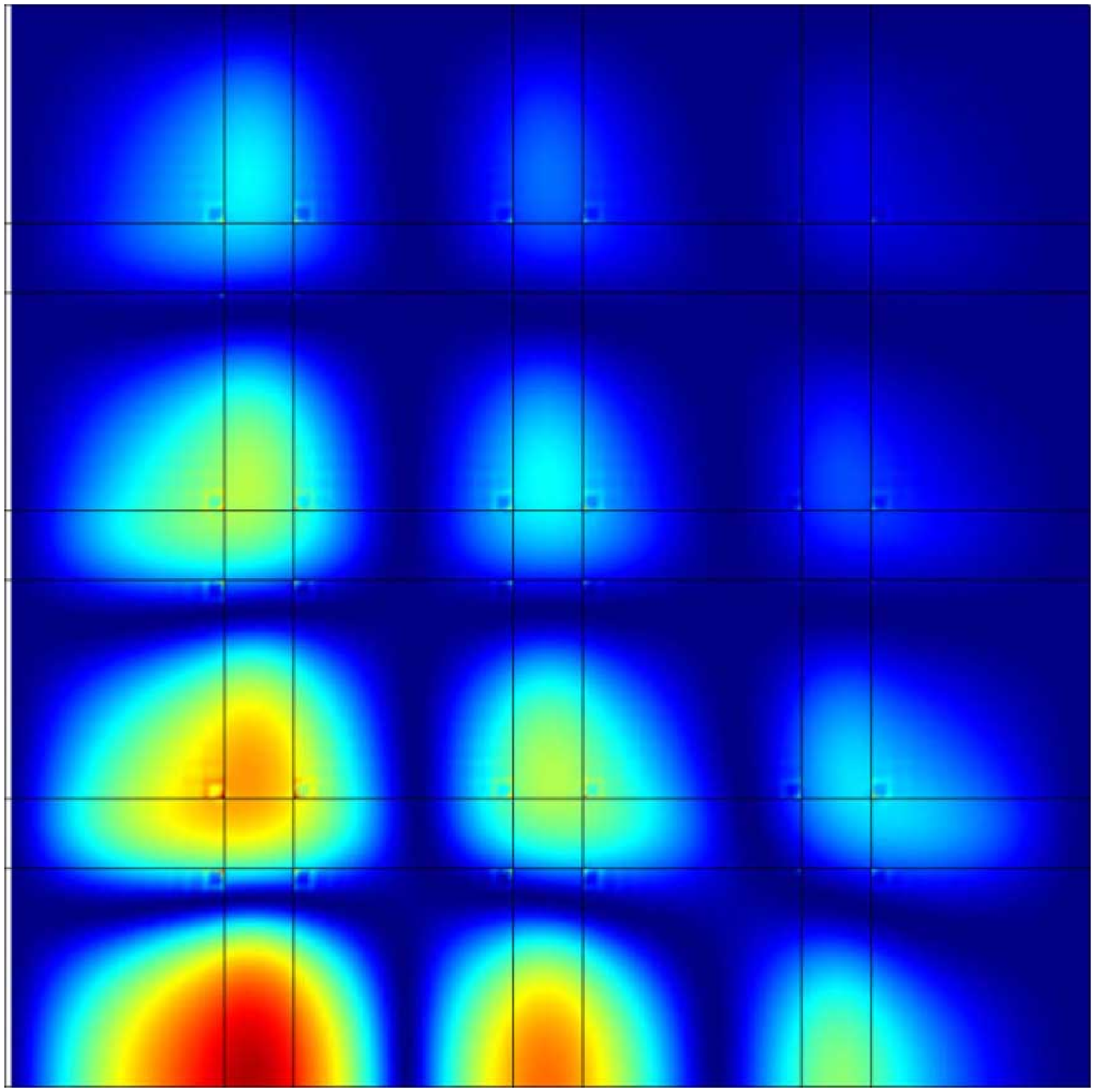}}\\
\\
c:$\:$ \mbox{\includegraphics[width=0.45\columnwidth]{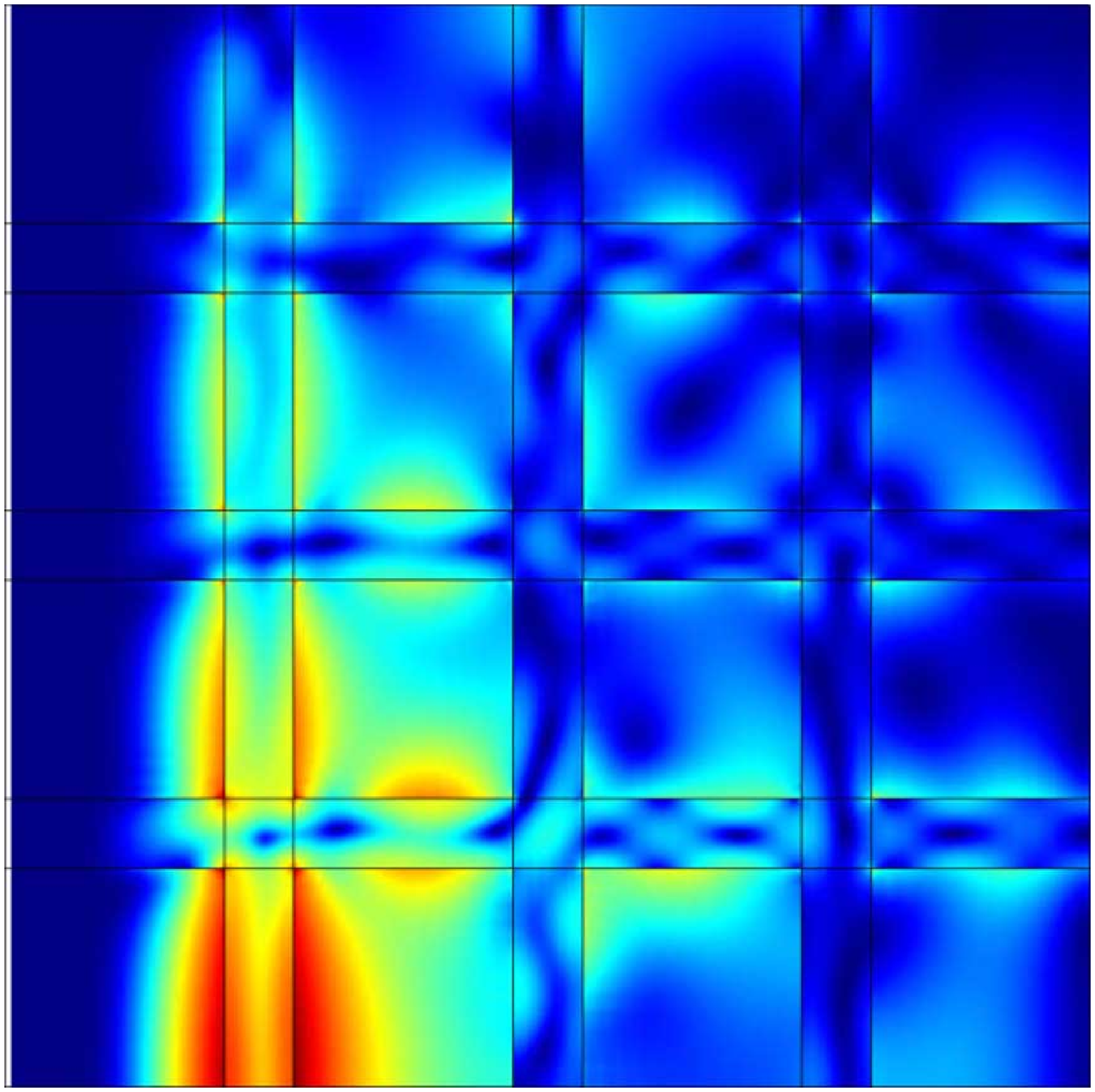}}
\end{tabular}
\caption{(a) Geometry (medial cross-section) of a alumina:air 3rd-order Bragg-cavity
resonator within a cylindrical metallic can (electric walls);
the can's interior surfaces are represented by a solid black line;
its interior diameter equals its interior height (and thus
this black line takes the form of a square);
the horizontal and vertical grey (or pink --in color reproduction)
stripes denote cylindrical plates and barrels, respectively, of alumina;
white squares correspond to regions of free-space (either air or vacuum);
the vertical arrow indicates the resonator's axis of rotational symmetry;
the dashed horizontal line (cf.~M1 in Fig.~\ref{fig:generic_resonator}) denotes a
plane of mirror symmetry, on which an electric or magnetic wall is imposed.
(b) False-color plot of the (logarithmic) electric-field intensity
$|\textbf{E}|^2$ for a zeroth-azimuthal-mode-order ($M = 0$) mode at
8.0873 GHz, localized towards the resonator's center (bottom left in figure);
(b) the same but for a sixth-azimuthal-mode-order ($M = 6$) mode at
20.0267 GHz, strongly localized in the radial directions but less so
in the vertical direction.}
\label{fig:BraggCavity}
\end{figure}
Through such a PDE-solver,
the method described in sections \ref{sec:Method} through \ref{sec:RadLoss} can be applied
to axisymmetric dielectric resonators of arbitrarily complex medial cross-section, the
only requirements being that each such cross-section is (i) bounded (either externally by an
enclosure or internally as an excluded region, or both) by metallic walls and (ii) decomposable
into definable regions of uniform dielectric.
This ability to cope with structural complexity is exemplified here in a modest way through the simulation
of a 3rd-order Bragg-cavity alumina:air microwave resonator whose geometry is shown in
Fig.~\ref{fig:BraggCavity}(a).
This resonator's model geometry was generated straightforwardly through a script written in MATLAB.
The resultant FEM mesh in COMSOL comprised 4356 base-mesh elements, with 53067 degrees
of freedom (DOF), corresponding to 12 edge vertices per $\lambda/4$
interval of air [\emph{i.e.}, across each white square in Fig.~\ref{fig:BraggCavity}(a)],
and 6 vertices per $\lambda/4$ interval of alumina [\emph{i.e.}, across each grey/pink `strip',
\emph{ibid.}].
Figs. ~\ref{fig:BraggCavity}(b) and (c) display two different calculated modes that
this resonator supports.

\section{Determination of the permittivities of cryogenic sapphire\label{sec:PermDet}}
The author harnessed the method of simulation constructed in sections \ref{sec:Method} and \ref{sec:Postprocessing}
to extract an independent determination of the two dielectric constants
of pure (HEMEX \cite{crystalsystems06}) monocrystalline sapphire at liquid-helium temperature,
based on some existing experimental data \cite{marra05}.
\begin{figure}[h]
\centering
\begin{tabular}{@{}l}
a:$\:$ \mbox{\includegraphics[width=0.5\columnwidth]{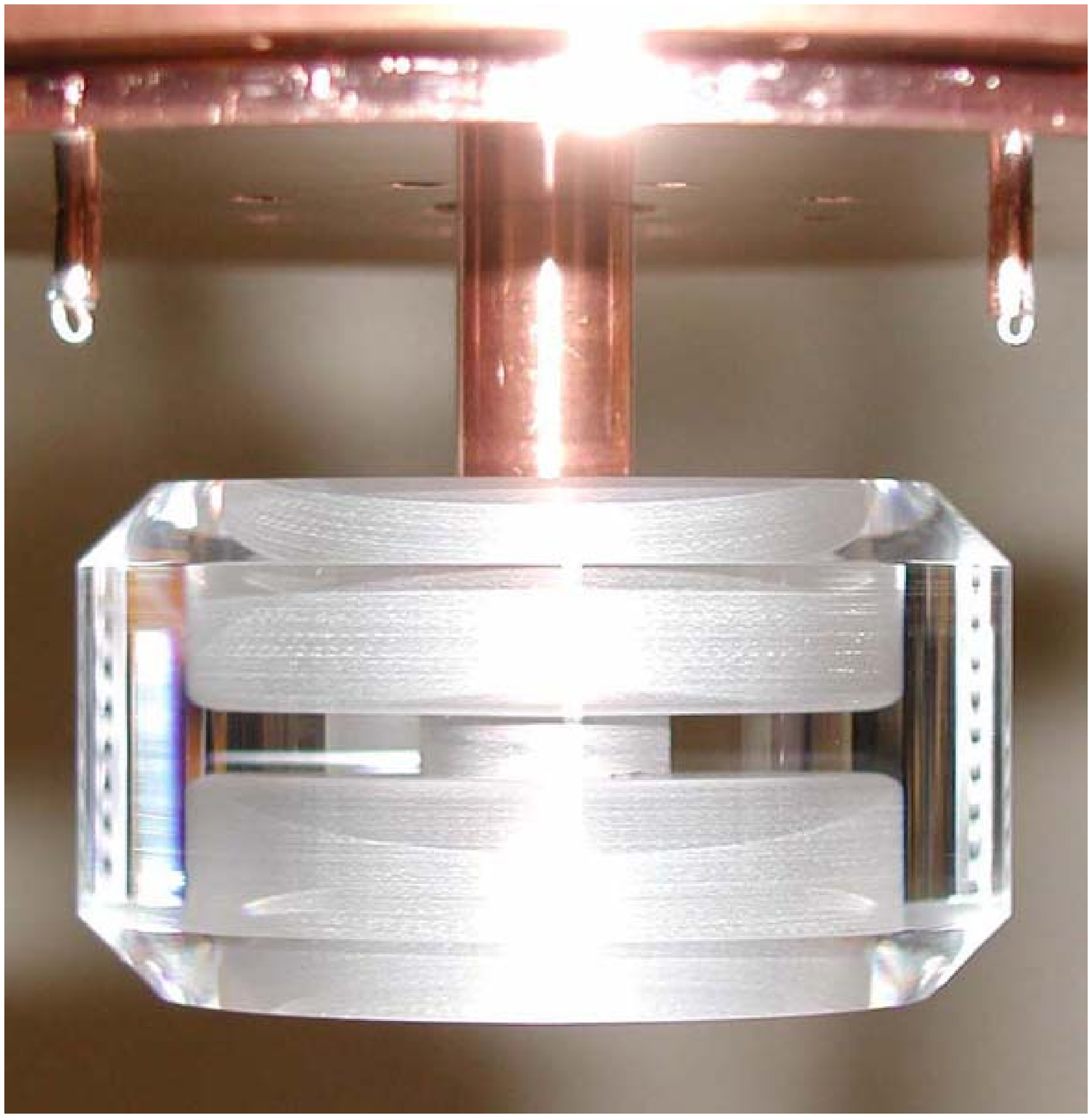}}\\
\\
\end{tabular}
\begin{tabular}{@{}l@{\quad}l}
b:$\:$ \mbox{\includegraphics[width=0.42\columnwidth]{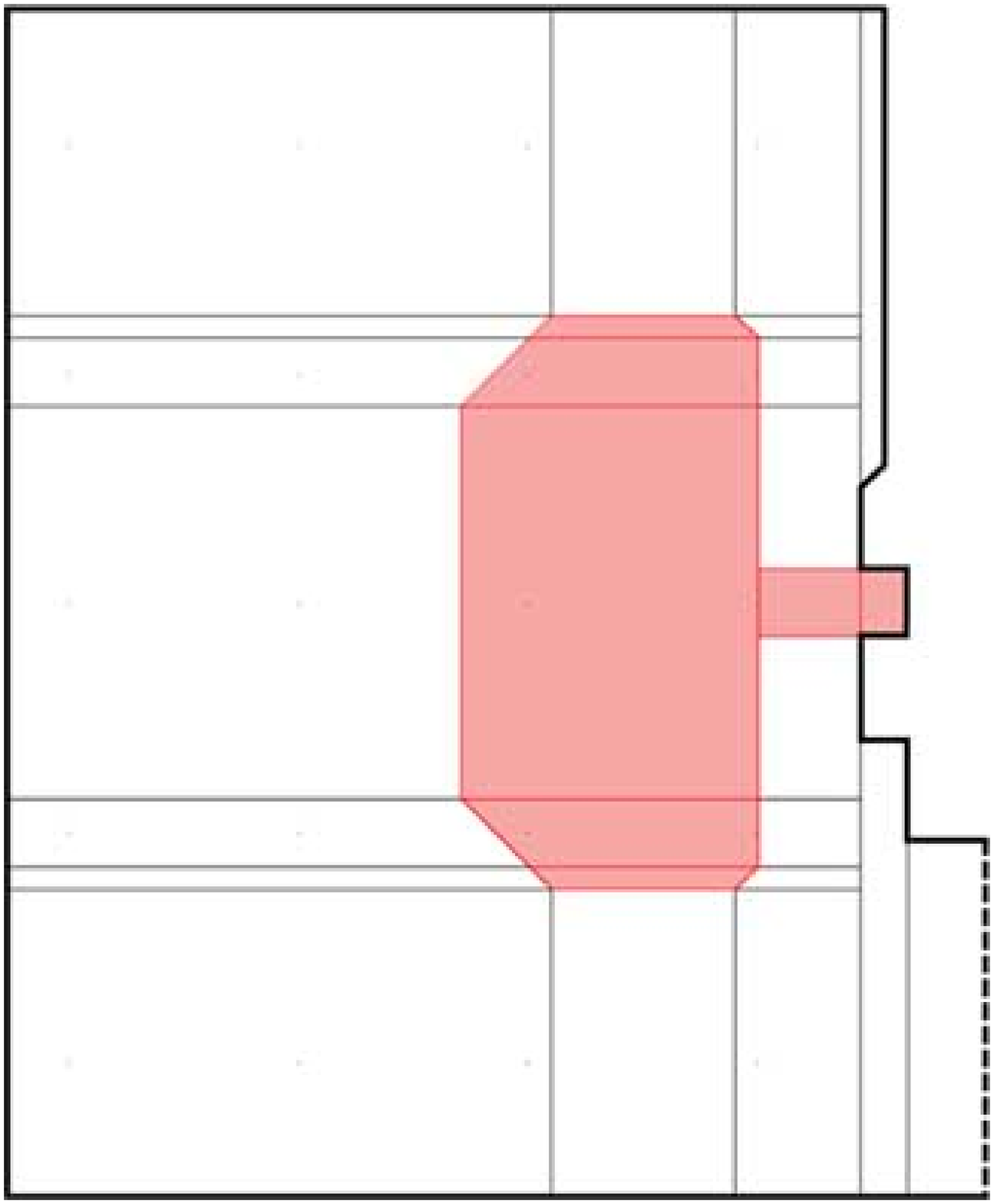}}
& c:$\:$ \mbox{\includegraphics[width=0.42\columnwidth]{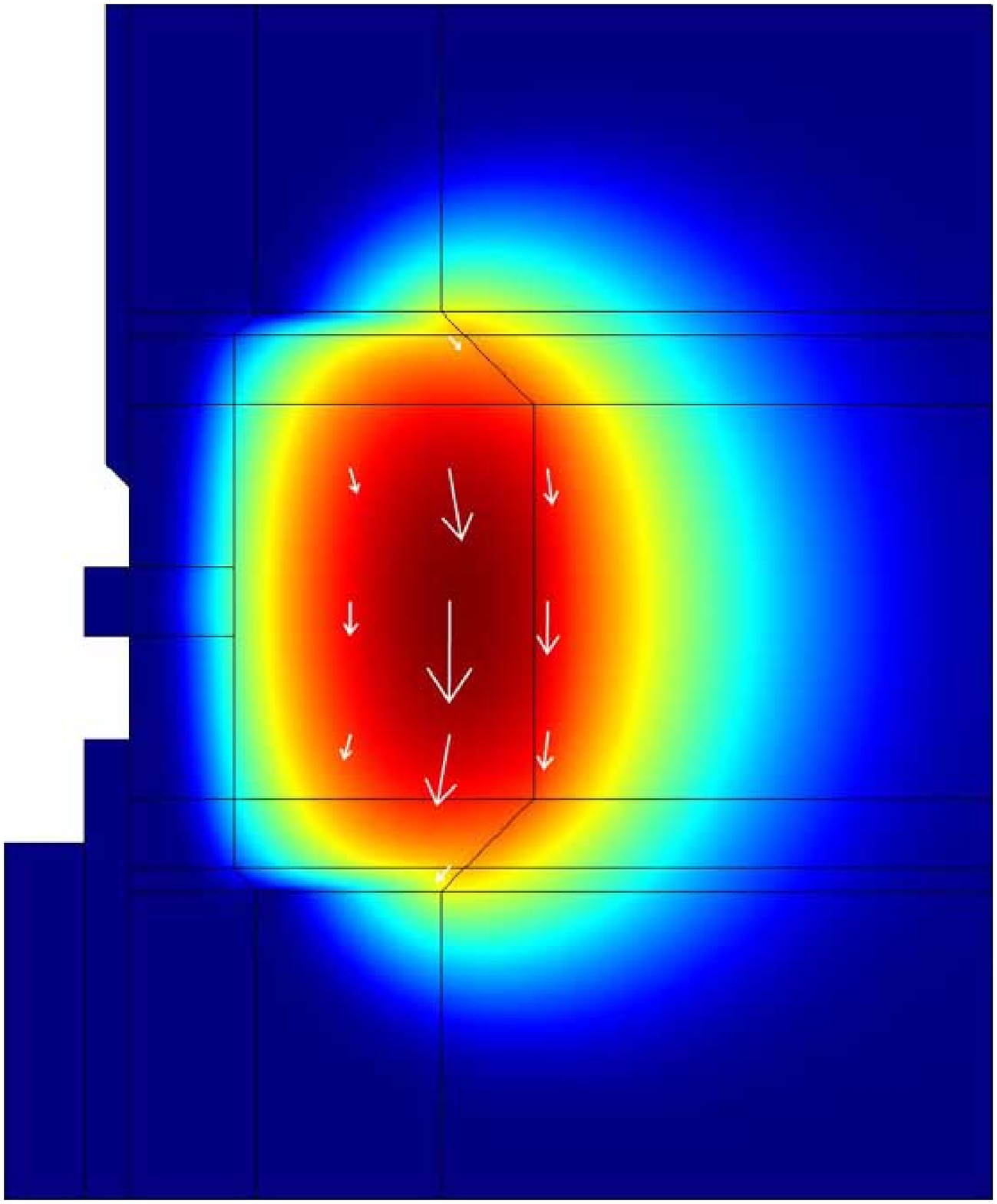}}\\
\end{tabular}
\caption{(a) Close-up of one of NPL's two (nominally identical)
Cs-fountain cryo-sapphire resonators, with its outer copper can removed.
The resonator's chamfered HEMEX sapphire ring has an outer diameter of $\sim$46.0~mm
and a axial height of $\sim$25.1~mm. This ring's integral interior `web', 3mm thick,
is oriented parallel to and centered (axially) on the ring's equatorial plane; the web is supported through
a central copper post, which is in turn connected [indirectly --through a thin, annular stainless steel
`shim' (not visible)] to the resonator's copper lid (onto which the removed can is secured).
Note that the sapphire's high refractive index falsely exaggerates [cf.~the true relative
dimensions shown in (b)]
the ring's internal diameter of $\sim$20.0 mm. Above the ring lie two loop probes for coupling,
electromagnetically,
to the resonator's operational whispering-gallery mode. [As finally configured, these probes
were withdrawn upwards several mm's closer to the lid and thus further (axially) from the ring].
(b) geometry of the resonator in medial cross-section; pink/grey indicates sapphire, white free space;
bounding these dielectric domains, and shown as thick solids lines, are copper surfaces belonging to the
resonator's can, lid and ring-supporting post; the resonator's cylindrical axis ($r$ or $x = 0$) is shown
as a dashed vertical line.
(c) false-color map (logarithmic scale) of the magnetic ($\textbf{H}$) field's squared
magnitude for the resonator's 11th-azimuthal-mode-order fundamental quasi-transverse-magnetic (N1$_{11}$ in
ref.~\cite{tobar01}'s notation) whispering-gallery mode at 9.146177 GHz (simulated),
as detailed on the 6th row of TABLE~\ref{tab:PermittivityFit}.
The white arrows indicate the magnitude and
direction of this mode's electric ($\textbf{E}$) field in the medial plane.}
\label{fig:NPLCsSapphRes}
\end{figure}
This data, as is listed in
the four right-most columns of TABLE~\ref{tab:PermittivityFit},
comprised\footnote{Though not listed in TABLE~\ref{tab:PermittivityFit},
the measured insertion loss (\emph{i.e.}~$S_{21}$ at line center)
for each resonance was also available.}: the centre frequencies,
FWHM widths, turnover temperatures, and `Kramers' splittings
for a set of 16 resonances, as measured on a (one of a pair of) cryogenic
sapphire resonator(s), as shown, without its enclosing can,
in Fig.~\ref{fig:NPLCsSapphRes}(a). Only two resonances out of this set (\emph{viz.}~N1$_{11}$
and S2$_9$) had hitherto been identified --via MAFIA \cite{mafia,oxborrow01} simulations\footnote{The
as-measured resonator was developed as part of a local `flywheel' oscillator for supplying
NPL's Cs-fountain(s) with an ultra-frequency-stable 9.1926 ... GHz reference, with the
resonator operating on (as it turned out) the S2$_9$ WG mode.}.
\begin{table}[t]
\caption{\label{tab:PermittivityFit}NPL's cryogenic sapphire
resonator: simulated and experimental WG modes compared}
\begin{minipage}{\linewidth}
\renewcommand{\thefootnote}{thempfootnote}
\begin{tabular}{l@{\quad}l@{\quad}l@{\quad}l@{\quad}l@{\quad}l@{\quad}l@{\quad}l}
\hline
Simulated   &Simul.       &Simul.       &Mode      &Experi-     &Exper.  &Exper.   &Exper.\\
minus       &perp.     &para.       &ID\footnote{the nomenclature of ref.~\cite{tobar01} is used for this column.}
                                                    &mental    &width\footnote{full width half maximum (-3 dB)}
                                                                        &turn-    &Kram.\footnote{the difference
in frequency between the orthogonal pair of standing-wave resonances (somewhat akin to a
`Kramers doublet' in atomic physics) associated with each WG model; the experimental parameters
stated in other columns correspond to the strongest resonance (greatest $S_{21}$ at line center)
of the pair.} \\
experim.    &filing      &filling      &          &freq.       &        &over   &split.\\
frequency   &factor      &factor       &          &            &        &temp.        &\\
$[$MHz$]$   &            &             &          &$[$GHz$]$   &$[$Hz$]$&$[$K$]$ &$[$Hz$]$\\
\hline
 -0.451     &0.860      &0.090      &S2$_6$      &6.954664    &285     &        &780\\
 -0.945     &0.930      &0.028      &S2$_7$      &7.696176    &82.5    &$<4.2$     &158\\
 0.881      &0.453      &0.517      &S4$_6$      &8.430800    &        &        &\\
 -1.538     &0.951      &0.014      &S2$_8$      &8.449908    &44.5    &$<4.2$     &418\\
 -0.412     &0.674      &0.299      &N2$_8$      &9.037458    &        &$4.8$     &\\
 -2.208     &0.071      &0.917      &N1$_{11}$   &9.148385    &9       &$5.0$     &57\\
 -1.916     &0.960      &0.009      &S2$_9$      &9.204722    &15.5    &$<4.2$     &88\\
 0.498      &0.251      &0.733      &S1$_{10}$   &9.267650    &12      &$5.2$     &180\\
 1.055      &0.287      &0.685      &N4$_8$      &9.421207    &80      &$5.0$     &\\
 -0.177     &0.437      &0.543      &S3$_8$      &9.800335    &84      &$4.8$     &1850\\
 0.358      &0.223      &0.763      &S1$_{11}$   &9.901866    &10      &$5.0$     &160\\
 -2.269     &0.965      &0.007      &S2$_{10}$   &9.957880    &24      &$<4.2$     &\\
 1.32       &0.730      &0.246      &S4$_8$      &10.27242    &153     &$5.0$     &\\
 0.19       &0.200      &0.787      &S1$_{12}$   &10.53863    &9.5     &$4.9$     &24\\
 0.00       &0.181      &0.808      &S1$_{13}$   &11.17728    &24.5    &$4.9$     &42\\
 4.13       &0.972      &0.006      &S2$_{12}$   &11.44918    &10      &$5.2$     &\\
\hline
\end{tabular}
\end{minipage}
\end{table}
This cryo-sapphire resonator's complete, detailed model geometry, as shown in Fig.~\ref{fig:NPLCsSapphRes}(b),
was coded into a MATLAB script. This script contained, for example, the dimensional parameters
specifying the form of the sapphire ring's (large) external and (smaller) internal chamfers.
The model geometry took
into account the shrinkages of the resonator's constituent materials from room temperature (293~K)
down to liquid-helium temperature (4.2~K).
The two cryo-shrinkages of sapphire (a uniaxial crystal) were calculated
by integrating up \cite{langham01} the linear-thermal-expansion data stated in Table~4
of ref.~\cite{white83} (identical to that stated in TABLE~1 of ref.~\cite{white93}):
$(1.0 - 7.21 \times 10^{-4})$ and $(1.0 - 5.99 \times 10^{-4})$ for directions
parallel and perpendicular to sapphire's c-axis, respectively. The cryo-shrinkage
of (isotropic) copper was taken directly from Table~F at the back of
ref.~\cite{white79}\footnote{Ref.~\cite{white80} provides linear-thermal-expansion
data for copper as a function of temperature --useful for
design purposes.}: $(1.0 - 3.26 \times 10^{-3})$.
The values of sapphire's two dielectric constants were initially
set equal to those specified in ref.~\cite{krupka99a}:
$\epsilon_{\perp} = 9.2725$ and  $\epsilon_{\parallel} = 11.3486$\footnote{These
values are consistent with $\epsilon_{\perp} = 9.272$ and $\epsilon_{\parallel} = 11.349$,
as stated in ref.~\cite{wolf04}.}.
\begin{figure}[b]
\centering
\mbox{\includegraphics[width=0.95\columnwidth]{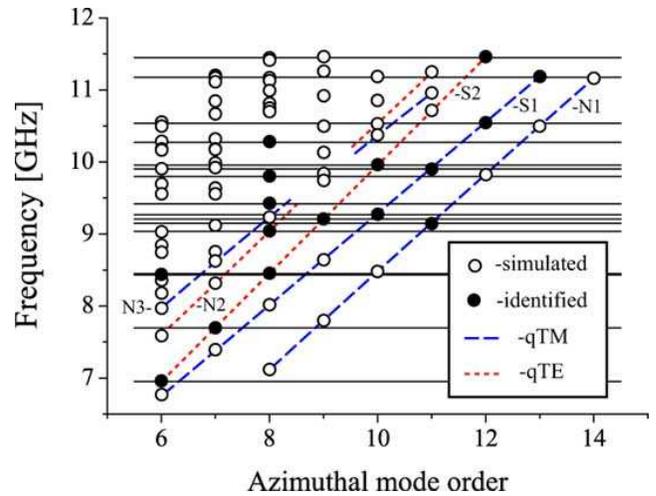}}
\caption{Plot used to identify experimentally measured with simulated WG modes. Solid
horizontal lines (16 in total) indicate the center frequencies of the former. Solid circles
indicate the identification of a simulated mode with an experimental one (the difference
in their frequencies corresponds to much less than a circle's radius in all cases);
hollow circles indicate simulated modes that were not identified with any experimentally
measured one. Quasi-transverse-magnetic (q-TM) and quasi-transverse-electric (q-TE) WG modes
of the same family are joined by (blue-)dashed and (red-)dotted lines respectively; a few of the
lowest-lying mode families are labeled using standard notation \cite{tobar01}.}
\label{fig:ExpvSimCsRes}
\end{figure}
Fig.~\ref{fig:NPLCsSapphRes}(b)'s geometry was meshed with quadrilaterals over the medial half-plane, with
8944 elements in its base mesh, and with DOF~$= 108555$. [These quadrilaterals followed the sloping
chamfers of the resonator by taking the shape of trapezoids.] For a given azimuthal mode order $M$, the
calculation of the lowest 16 eigenmodes took around 3 minutes on the author's office PC (as previously
specified). Fig.~\ref{fig:NPLCsSapphRes}(c) shows the form of the resonator's N1$_{11}$ whispering-gallery mode,
corresponding to the 6th row of TABLE~\ref{tab:PermittivityFit}. Filling factors were then
calculated to quantify each frequency's sensitivity to changes in the sapphire's two dielectric constants
($\epsilon_{\parallel}$ and $\epsilon_{\perp}$). The author identified each of the 16 experimental resonances with
a particular simulated WG mode, aiming to minimize the residual (simulated-minus-measured)
frequency difference (\emph{i.e.}~the sum $\chi^2$ variance over the left-most column in
TABLE~\ref{tab:PermittivityFit}), whilst requiring that the
other measured attributes (\emph{e.g.}~insertion loss, linewidth) of the
resonances identified to the same `family' of WG modes (\emph{e.g.}~S1 or N2)
varied smoothly with the azimuthal mode order $M$.
With the identifications of the experimental modes `locked' as
per the 4th column of TABLE~\ref{tab:PermittivityFit}, the model's two sapphire
dielectric constants were adjusted from their initial values to minimize
$\chi^2$ (`least squares'). The resultant best-fit values were:
\begin{eqnarray}
\label{eq:fitted_e_perp}
\epsilon_\perp & = & 9.285 \,\,(\pm 0.010);\\
\label{eq:fitted_e_para}
\epsilon_\parallel  & = &11.366 \,\,(\pm 0.010).
\end{eqnarray}
With the two dielectric constants set to these values, the
WG modes were recalculated (an in-principle superfluous check);
the first three columns in TABLE~\ref{tab:PermittivityFit} result
from this recalculation (the filling factors hardly changed from
their original fitted values).
Pending the construction of a more detailed and precise error budget,
the provisional $\pm 0.010$ uncertainty assigned to the values of both
dielectric constants reflects their observed shifts upon refitting with
a few `problematic' experimental modes identified with different
simulated ones
\footnote{A few of
the experimentally measured modes had a number simulated modes in close proximity to their center
frequencies; note that the
centers of several circles lie close to certain horizontal lines in see Fig.~\ref{fig:ExpvSimCsRes}.
After considering all available pieces of experimental information (\emph{viz.} linewidth,
insertion loss, and turnover temperature), doubt still remained as to the correct identifications
for some of them.
The 4th column of TABLE~\ref{tab:PermittivityFit}
represents the most likely, but not the only conceivable, set.
Though a `prettier' (and perhaps, even, more accurate) determination could have
been presented by dropping these problematic modes/identifications
from the least-squares fit, the author --given the purposes of this paper-- elected
to fit all 16 experimentally measured resonance, keeping the generic problem
of mode identification to the fore.]}.
Compared to these identification-related shifts, the systematic errors associated with a finite
meshing density --as analyzed quantitatively in ref~\cite{santiago94}),
or the experimental uncertainties associated
with the resonator's geometric shape (particularly the diameter and height of the sapphire ring)
were quite negligible. The specified (\emph{i.e.}~contracted) tolerance on the alignment of the sapphire
crystal's c-axis with respect to the geometric axis of the `cored' cylinder from which the two
rings were cut (through orientation-preserving methods) was only $< 0.5$ degrees.
Though the effects of crystal misalignment cannot be modeled quantitatively with the method presented
in this paper, for which axial symmetry is a requirement, it can be estimated that, given the  $\sim$two-parts-in-ten
contrast between the sapphire's parallel and transverse permittivities, such a misalignment
should make a significant/dominant contribution to the ~1-part-in-a-thousand residuals
between the simulated and measured center frequencies (and thus the determination of
$\epsilon_\perp$ and $\epsilon_\parallel$.) Even with azimuthal mode orders
of $M \sim 10$, as is the case here, the narrowness of the measured WG resonances'
Kramers splittings (listed in the right-most column of TABLE~\ref{tab:PermittivityFit},
and generally less than 1 part per million relative to the absolute frequency)
would indicate a much higher degree of rotational invariance, however.
Though noting that center-frequency residuals of a few
parts per thousand are not untypical for FEM-based simulations of WG modes \cite{santiago94},
the author has yet to reconcile, convincingly, the residuals with their cause(s) --as would
be required to construct a more detailed error budget.

\section{Conclusion}
This paper demonstrates, through the explicit statement of weak-form expressions and
boundary constraints, how a commercial (FEM-based) PDE-solver can be configured
to simulate, quickly and to high accuracy, the whispering-gallery modes of axisymmetric
dielectric resonators on standard computer hardware.
The source codes/configuration scripts used to implement the simulations presented
in section \ref{sec:ExampleApplications} of this paper are freely available
from the author.


%


\section*{Acknowledgment}
The author thanks Anthony Laporte and Dominique Cros at the IRCOM, Limoges, France,
for unexpectedly supplying him with an independent (and corroborating) set of simulated
resonance (center) frequencies for the chamfered cryogenic-sapphire resonator considered in
section~\ref{sec:PermDet} --as produced via their own (2D) electromagnetic software.
The author also thanks two NPL colleagues: Giuseppe Marra,
for supplying and/or verifying a considerable fraction of the
experimental data presented within Table~\ref{tab:PermittivityFit};
and Louise Wright, for a detailed review of the manuscript prior to submission.
%




\bstctlcite{IEEE:BSTcontrol}



\bibliographystyle{IEEEtran}
%
%
%

%

\begin{biography}[{\includegraphics[width=1in,height=1.25in,clip,keepaspectratio]{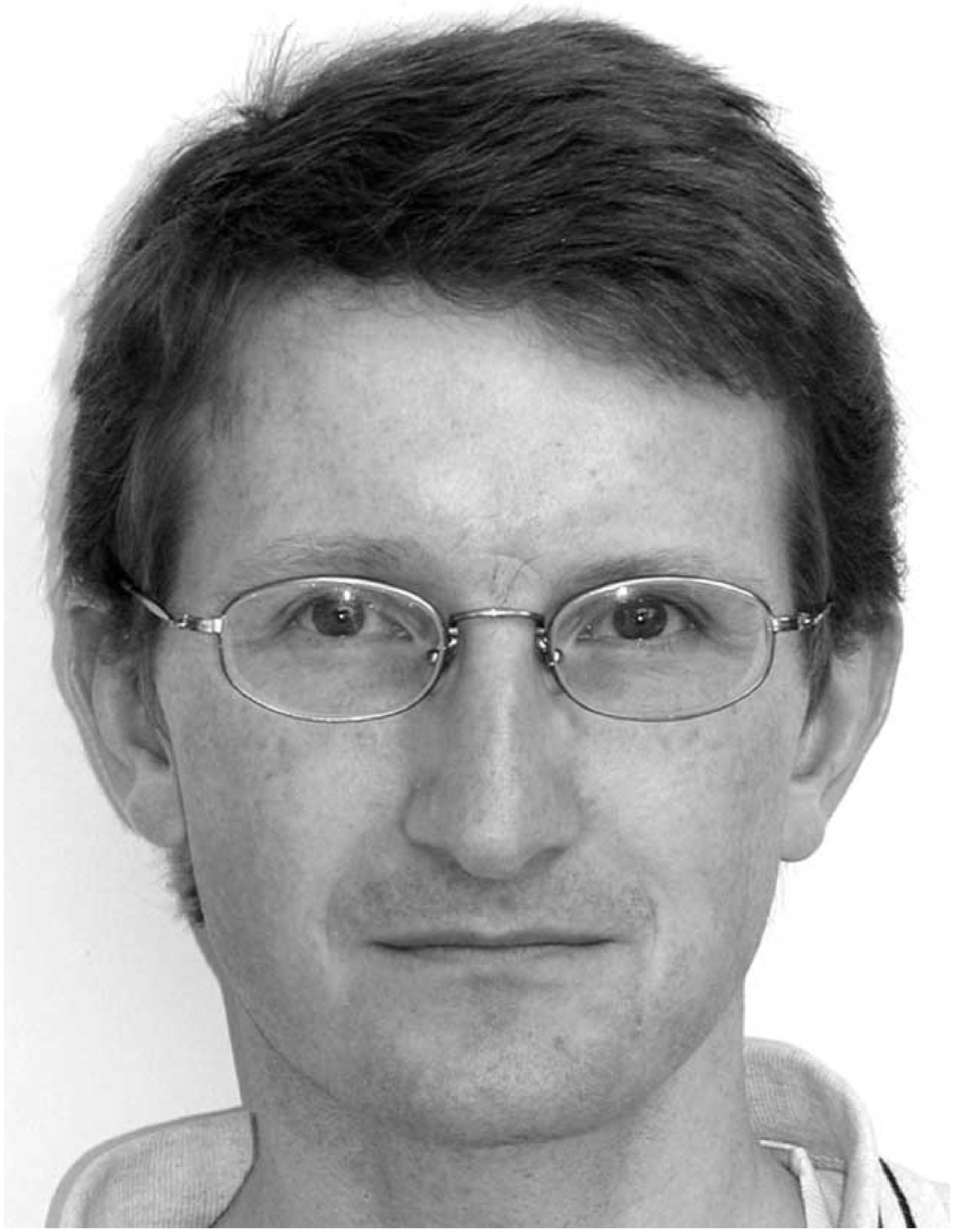}}]{Mark Oxborrow}
was born near Salisbury, England, in 1967. He received a B.A. in physics from the University of Oxford in 1988, and
a Ph.D. in theoretical condensed-matter physics from Cornell University, Ithaca, NY, in 1993; his thesis topic concerned
random-tiling models of quasicrystals.
During subsequent postdoctoral appointments at both the Niels Bohr Institute in Copenhagen and back
at the University of Oxford, he investigated acoustic analogues of quantum wave-chaos.
In 1998, he joined the UK's National Physical Laboratory; his research there to date has included
the design and construction of ultra-frequency-stable microwave and optical oscillators, the development
of single-photon sources, and the applications of carbon nanotubes to metrology.
\end{biography}


\newpage
.
\newpage






\appendices

\section{Configuration of COMSOL Multiphysics for simulating axisymmetric
dielectric resonators: explicit weak-form expressions}
It is here explained, in some detail, how to set up a dielectric-resonator
simulation in COMSOL Multiphysics \cite{COMSOL} --from scratch.
These explanations should also be helpful to anyone wishing to modify one
of the author's existing models --as incarnated in an .MPH file.
At least in the first instance,
it is recommended that the following instructions be meticulously
adhered to --lest one stray from a tried-and-tested path.
And it is suggested that the reader work through them
with COMSOL Multiphysics open and running on his/her desktop. All
menu items, expression names and variables associated with the program are
displayed in \verb"typed text" font.
A good deal of supplementary information can be found in the documentation
supplied with COMSOL Multiphysics itself; the author found the following
chapters therein to be the most useful/relevant:
\emph{`PDE Modes for Equation-Based Modeling'},
\emph{`The Weak Form'},
and \emph{`COMSOL Multiphysics Scripting'}.
Upon reading these chapters, one might be left with the impression
that COMSOL is simply not sufficiently flexible to embrace the
task in hand (\emph{i.e.}~to implement sections \ref{sec:Method}
through \ref{sec:RadLoss} of this article explicitly); the following instructions demonstrate
how COMSOL Multiphysics can, despite these first impressions, and most
straight-forwardly, be so configured
to implement the 2D simulation of isotropic dielectric resonators.
From the beginning then:
\begin{table}[t]
\caption{\label{tab:COMSOLConstants}COMSOL Constants --included in weak-form expressions}
\begin{minipage}{\linewidth}
\renewcommand{\thefootnote}{thempfootnote}
\begin{tabular}{@{}l@{\quad}l@{\quad}l@{}}
\hline
Name                        &Expression             &Description [unit]\\
                            &(= value)              &\\
\hline
\verb"c"                    &299792458              &speed of light (exact) [m/s]\\
\verb"cbar"                 &\verb"2*pi/c"          &frequency constant [s/m]\\
                            &( = 2.095845e-8)       &\\
\verb"cbar2"                   &\verb"4*pi^2/c^2"      &frequency constant [s$^2$/m$^2$]\\
                            &( = 4.392566e-16)      &\\
\verb"alpha"                &1.0                    &penalty-term coefficient\\
\verb"M"                    &9                      &azimuthal mode order\\
\verb"e0"                   &1.0                    &relative permittivity\\
                            &                       &of free-space\\
\verb"e1"                   &\verb"n_AlGaAs^2"      &relative permittivity of\\
                            &( = 11.2896)           &isotropic\_dielectric\_1\\
\verb"e2"                   &1.0                    &same but of\\
                            &                       &isotropic dielec.\_2\\
\verb"e3"                   &1.0                    &etc.\\
\verb"eperp0"               &1.0                    &relative permittivity\\
                            &                       &of free-space\\
                            &                       &in directions transverse\\
                            &                       &to cylindrical axis\\
\verb"epara0"               &1.0                    &same but in direction\\
                            &                       &parallel to cylindrical axis\\
\verb"eperp1"               &9.2725                 &relative permittivity of\\
                            &                       &uniaxial\_dielectric\_1\\
                            &                       &in directions transverse\\
                            &                       &to cylindrical axis\\
\verb"epara1"               &11.3486                &same but in direction\\
                            &                       &parallel to cylindrical axis\\
\verb"epara2"               &1.0                    &relative permittivity of\\
                            &                       &uniaxial\_dielectric\_2\\
                            &                       &transverse to cyl. axis\\
\verb"eperp2"               &1.0                    &same but parallel to\\
                            &                       &cylindrical axis\\
\verb"eperp3"               &1.0                    &etc.\\
\verb"epara3"               &1.0                    &etc.\\
\verb"e_293K_alumina"       &9.8                    &relative permittivity of\\
                            &                       &alumina at room temperature\\
\verb"epe_4K_sap_UWA"       &9.2725                 &UWA values for cryogenic\\
                            &                       &HEMEX sapphire\\
\verb"epa_4K_sap_UWA"       &11.3486                &\\
\verb"epe_293K_sap"         &9.407                  &nominal room temperature\\
                            &                       &values for same\\
\verb"epa_293K_sap"         &11.62                  &\\
\verb"epe_4K_sap_NPL"       &9.2848                 &Values fitted to\\
                            &                       &NPL Cs-fountain\\
                            &                       &HEMEX resonator\\
\verb"epa_4K_sap_NPL"       &11.3660                &\\
\verb"n_silica"             &1.4457                 &refractive index of\\
                            &                       &thermally grown\\
                            &                       &silica (Fig B.2, p. 172 of\\
                            &                       &ref.~\cite{kippenberg04})\\
\verb"n_AlGaAs"             &3.36                   &average refractive index of\\
                            &                       &GaAs and AlGaAs layers\\
                            &                       &(p.~172 of ref.~\cite{srinivasan06})\\
\verb"mf"                   &2.374616e14            &match frequency\\
\verb"ttgH"                 &1                      &toggle\\
\verb"ttgE"                 &0                      &toggle\\
\verb"mix_ang"              &45                     &electric-magnetic mixing\\
                            &                       &angle (in degrees)\\
\verb"cMW"                  &\verb"sin(mix_ang"     &\\
                            &\verb" * pi /180)"     &magnetic-walled-ness\\
                            &(= 0.707107)          &\\
\verb"cEW"                  &\verb"cos(mix_ang"     &\\
                            &\verb" * pi /180)"     &electric-walled-ness\\
                            &(= 0.707107)          &\\
\verb"tngM"                 &1                      &\\
\verb"tngE"                 &0                      &\\
\hline
\end{tabular}
\end{minipage}
\end{table}
\subsection{Setting up --fundamentals}
Get COMSOL Multiphysics up and running.
Access the \verb"Model Navigator" panel via \verb"File" $\Rightarrow$ \verb"New ..." and
select the \verb"New" tab if not already selected.

(a) Select `\verb"2D"' from the \verb"Space dimension:" drop-down menu [note: do {\it not} choose
`\verb"Axial symmetric (2D)"'].

(b) Browse to and select `\verb"COMSOL Multiphysics" $\Rightarrow$ \verb"PDE Modes"
$\Rightarrow$ \verb"Weak Form, Subdomain"' from the \verb"Application Mode" navigator.

(c) Type (verbatim) `\verb"Hrad" \verb"Hazi" \verb"Haxi"' into the \verb"Dependent variables:" text field.
These three variables are the radial, azimuthal and axial components of the magnetic field strength, respectively;
all three are dependent on (\emph{i.e.}~are functions of) the Cartesian coordinates for the COMSOL simulation's
2D space, namely \verb"x" (horizontal on the screen) and \verb"y" (vertical) --both in units of metres [m].
The coordinate names `\verb"x"' and `\verb"y"' are already fixed by COMSOL
(\emph{i.e.}~they are reserved symbols) and need not be explicitly entered
(in COMSOL terminology, \verb"x" and \verb"y" are `geometric variables').

(d) For the \verb"Application mode name:" (default \verb"u") one can type in anything one likes.

(e) Select `\verb"Lagrange - Quadratic"' from the \verb"Element:" drop-down menu.
[This choice is proven to work.]

\subsection{Constants}
All of the various constants (\emph{i.e.} independent of \verb"x" or \verb"y") included
within the weak-form expressions given below are defined and described in TABLE~\ref{tab:COMSOLConstants}.
The equivalent of this table needs to be typed (or loaded) into COMSOL's \verb"Options" $\Rightarrow$ \verb"Constants ...".
Each \verb"Expression" thus \verb"Value" therein [except those for \verb"e0", \verb"eperp0", \verb"epara0" --which
define the (unit) relative permittivity of free-space], can be user-varied.
But every \verb"Name" should be entered verbatim; \emph{i.e.}, each constant
must be named exactly as it appears in the expressions that subsequently include it.

\subsection{Expressions (for Postprocessing)}
The post-processing of the calculated magnetic-field strength
(as a function of position) for each solved eigenfunction is
facilitated through the various definitions presented here.
\subsubsection{Scalar expressions}
The equivalent of TABLE~\ref{tab:COMSOLscalarexpress}
(or some subset thereof) needs to be typed into
COMSOL's \verb"Options" $\Rightarrow$
\verb"Expression ..." $\Rightarrow$ \verb"Scalar expressions ...".
\begin{table}[t]
\caption{\label{tab:COMSOLscalarexpress}COMSOL Scalar Expressions --for postprocessing}
\begin{minipage}{\linewidth}
\renewcommand{\thefootnote}{thempfootnote}
\begin{tabular}{@{}l@{\quad}l@{\quad}l@{}}
\hline
Name                    &Expression\\
                        &(Description)\\
\hline
\verb"DivH"             &\verb"(Hrad-Hazi*M+(Haxiy+Hradx)*x)/x"\\
                        &(divergence of magnetic field --should be zero!)\\
\verb"MagEnDens"        &\verb"Hrad*Hrad+Hazi*Hazi+Haxi*Haxi"\\
                        &(magnetic energy density)\\
\verb"Drad"             &\verb"(Haxi*M-Haziy*x)/x"\\
                        &(radial component of electric displacement)\\
\verb"Dazi"             &\verb"-Haxix+Hrady"\\
                        &(azimuthal component of electric displacement)\\
\verb"Daxi"             &\verb"(Hazi-Hrad*M+Hazix*x)/x"\\
                        &(axial component of electric displacement)\\
\verb"Erad"             &\verb"Drad/eperp"\\
                        &(radial component of electric field strength)\\
\verb"Eazi"             &\verb"Dazi/eperp"\\
                        &(azimuthal component of electric field strength)\\
\verb"Eaxi"             &\verb"Daxi/epara"\\
                        &(axial component of electric field strength)\\
\verb"ElecMagSqrd"      &\verb"Erad*Erad+Eazi*Eazi+Eaxi*Eaxi"\\
                        &(electric field strength magnitude squared)\\
\verb"ElecEnDens"       &\verb"Erad*Drad+Eazi*Dazi+Eaxi*Daxi"\\
                        &(electric energy density)\\
\verb"AbsMagEnDens"     &\verb"abs(Hrad)^2+abs(Hazi)^2"\\
                        &\verb"+abs(Haxi)^2"\\
                        &(absolute magnitude energy density)\\
\verb"MagNrmlHSqrd"     &\verb"2*pi*x*abs(Haxi*ny"\\
                        &\verb"+Hrad*nx)^2"\\
                        &(magnitude normal mag.~field strength squared)\\
\verb"MagTngHSqrd"      &\verb"2*pi*x*(1*abs(Hazi)^2"\\
                        &\verb"+1*abs(Haxi*nx-Hrad*ny)^2)"\\
                        &(magnitude tangential magnetic field squared)\\
\verb"AbsElecSqrd"      &\verb"abs(Erad)^2+abs(Eazi)^2"\\
                        &\verb"+abs(Eaxi)^2"\\
                        &(absolute electric field squared)\\
\hline
\end{tabular}
\end{minipage}
\end{table}
\subsubsection{Subdomain expressions}
The functionality of \verb"Subdomain expressions" is required for generating post-processed fields,
like the electric field strength $\textbf{E}$ --as per the 6th, 7th and 8th entries in
TABLE~\ref{tab:COMSOLscalarexpress}. Those constants associated with each such field's
definitions, like (in the case of $\textbf{E}$) the relative permittivities \verb"epara"
and \verb"eperp", \emph{vary} from one subdomain within the medial half place to another.
The variation of these subdomain-dependent `constants' is represented through
\verb"Options" $\Rightarrow$ \verb"Expressions" $\Rightarrow$ \verb"Subdomain expressions";
therein, the \verb"Name" of each such variable is the same in each and every \verb"Subdomain" (as identified
by an integer), but its \verb"Expression" reflects the variable value in the selected \verb"Subdomain".
Thus, the \verb"Expression" for \verb"epara" in a \verb"Subdomain" corresponding to (cryogenic and
axisymmetrically oriented) sapphire would be \verb"epara1", with \verb"epara1" defined (globally)
as 11.3486 (or whatever) through TABLE~\ref{tab:COMSOLConstants}, whereas in a \verb"Subdomain"
corresponding to free space, the \verb"Expression" for \verb"epara" should be set to
\verb"1". Similarly (and more simply), the single \verb"Subdomain"-dependent variable \verb"erel" can be
used to represent the variation of relative permittivity within an axisymmetric resonator containing
solely isotropic dielectrics (incl.~free space).

\subsection{Weak-form expressions}
The simulation's defining weak-form expressions are set up through the
\verb"Physics" $\Rightarrow$ \verb"Subdomain Settings ..." control panel. On the left of this panel,
first select the \verb"Groups" tab. A \verb"New" \verb"Group" must be named and defined for
each dielectric within the resonator being simulated. The author chose to name these
dielectric \verb"Groups" `\verb"dielectric_0:vacuum"', `\verb"dielectric_1"', ..., `\verb"dielectric_n"', ... .
For each dielectric \verb"Group", (in general) \verb"dielectric_n" say, corresponding weak-form
expressions need to be entered into the \verb"weak terms" (\emph{i.e.}~three slots or text fields),
for expressions involving spatial derivatives, and also into
the \verb"dweak terms", for expressions involving temporal derivatives; these slots are
accessed through the \verb"weak" and \verb"dweak" tabs, respectively,
located on the right of the \verb"Subdomain Settings ..." control panel.
These terms govern the electromagnetic field in regions filled with the $n$-th dielectric,
as specified in the \verb"Constants ..." table introduced above.
No other fields on the right of \verb"Subdomain Settings ..." need(/should) be touched;
in particularly, don't monkey with the \verb"contr"-tabbed sub-panel. Note that it is imperative that
the \verb"Name" of each constant entered into \verb"Options" $\Rightarrow$ \verb"Constants ..."
above match (verbatim) its appearances within the expressions (below) that are entered into the
\verb"weak" and \verb"dweak" text fields here.
For each dielectric \verb"Group", two \verb"weak" and one \verb"dweak" terms are required:
(i) a `Laplacian' term (corresponding to the left most term on the left-hand side of equation \ref{eq:helmholtz}
and (ii) a `penalty' term, included to suppress spurious modes, corresponding to the middle
term of the same.
\subsubsection{Laplacian term [first weak-term slot]}:
The form of the Laplacian weak term,
$(\boldsymbol{\nabla} \pmb{\times} \tilde{\textbf{H}}^*) \frac{\cdot}{\boldsymbol{\epsilon}} \; (\boldsymbol{\nabla} \pmb{\times} \textbf{H})$,
here given for the 1st axisymmetric dielectric, is\footnote{
Note that, when typing (or `cutting-and-pasting') this and the following (d)weak-form expressions
into their slots, \emph{all spaces and new lines} must be eliminated from the whole expression within each
slot --otherwise COMSOL will reject the expression.}
\begin{equation}
\begin{tabular}{l}
\verb"((eperp1*(test(Hazi)*Hazi"\\
\verb"-M*(test(Hazi)*Hrad"\\
\verb"+Hazi*test(Hrad))"\\
\verb"+M^2*test(Hrad)*Hrad)"\\
\verb"+epara1*M^2*test(Haxi)*Haxi)/x"\\
\verb"+eperp1*(test(Hazix)*(Hazi-M*Hrad)"\\
\verb"+Hazix*(test(Hazi)-M*test(Hrad)))"\\
\verb"-epara1*M*(test(Haxi)*Haziy"\\
\verb"+Haxi*test(Haziy))"\\
\verb"+x*(eperp1*test(Hazix)*Hazix"\\
\verb"+epara1*((test(Haxix)"\\
\verb"-test(Hrady))*(Haxix-Hrady)"\\
\verb"+Haziy*test(Haziy)))"\\
\verb")/(epara1*eperp1)"
\end{tabular}
\end{equation}
where \verb"Hazix" denotes the partial derivative of \verb+Hazix+
with respect to the coordinate \verb"x", \verb"Hrady" the partial derivative of \verb"Hazi"
with respect to \verb"y", etc.; \verb"test(Hazi)" denotes the `test function' of \verb"Hazi", \emph{etc.}
Its equivalent for the 2nd axisymetric dielectric is obtained by replacing
\verb"eperp1" by \verb"eperp2" and \verb"epara1" by \verb"epara2",
and so forth for all other axisymetric dielectrics (should more be required).
The above expression can be significantly simplified for the (subdomain) \verb"Groups" corresponding
to isotropic dielectrics or free space ({\it viz.}~\verb"dielectric_0"); for computational efficiency,
it is recommended that these simplifications be implemented wherever possible.
The required form of the Laplacian \verb"weak term",
$[(\boldsymbol{\nabla} \pmb{\times} \tilde{\textbf{H}}^*) \cdot (\boldsymbol{\nabla} \pmb{\times} \textbf{H})] / {\epsilon_1} $,
for the 1st isotropic dielectric is given explicitly as
\begin{equation}
\label{eq:COMSOL_laplacianweak_isotrop}
\begin{tabular}{l}
\verb"((test(Hazi)*Hazi"\\
\verb"-M*(test(Hazi)*Hrad"\\
\verb"+Hazi*test(Hrad))"\\
\verb"+M^2*(test(Hrad)*Hrad"\\
\verb"+test(Haxi)*Haxi))/x"\\
\verb"+(test(Hazix)*(Hazi-M*Hrad)"\\
\verb"+Hazix*(test(Hazi)-M*test(Hrad)))"\\
\verb"-M*(test(Haxi)*Haziy"\\
\verb"+Haxi*test(Haziy))"\\
\verb"+x*(test(Hazix)*Hazix"\\
\verb"+((test(Haxix)"\\
\verb"-test(Hrady))*(Haxix-Hrady)"\\
\verb"+Haziy*test(Haziy))))/e1" ,
\end{tabular}
\end{equation}
where \verb"e1" is the material's dielectric constant (as appearing in TABLE~\ref{tab:COMSOLConstants}).
The Laplacian \verb"weak term" for the vacuum is the same with $\epsilon_1 \rightarrow 1$, and those
for other isotropic dielectrics are similarly obtained by swopping $\epsilon_1$ with $\epsilon_2$, $\epsilon_3$,
and so forth.
\subsubsection{Penalty (divergence-suppressing) term [second weak-term slot]}:
The form of the penalty \verb"weak term",
$\alpha (\boldsymbol{\nabla} \cdot \tilde{\textbf{H}}^*) \cdot (\boldsymbol{\nabla} \cdot \textbf{H})$,
the same for each subdomain \verb"Group", is
\begin{equation}
\begin{tabular}{l}
\label{eq:COMSOL_penaltyweak}
\verb"alpha*((test(Hrad)*Hrad"\\
\verb"-M*(test(Hazi)*Hrad"\\
\verb"+Hazi*test(Hrad))"\\
\verb"+M^2*test(Hazi)*Hazi)/x"\\
\verb"+(test(Haxiy)"\\
\verb"+test(Hradx))*(Hrad-M*Hazi)"\\
\verb"+(test(Hrad)-M*test(Hazi))"\\
\verb"*(Hradx+Haxiy)"\\
\verb"+x*(test(Hradx)"\\
\verb"+test(Haxiy))*(Hradx+Haxiy))"\\
\end{tabular}
\end{equation}
here, the coefficient \verb"alpha" (whose value is determined
through COMSOL's equivalent of TABLE~\ref{tab:COMSOLConstants})
controls the aggressiveness of the divergence suppression induced by this term.
The remaining, 3rd slot, should be zero-filled. [As a general rule,
unused weak-form slots should always be filled with zeroes --this applies to the the \verb"dweak term" slots
below.]

\subsubsection{Frequency term [first dweak-term slot]}
The form of the temporal-derivative/frequency (so-called `dweak') term
$\tilde{\textbf{H}}^* \cdot \partial^2\textbf{H}/{\partial^2 t}$,
common to all subdomain \verb"Groups",
is entered into the first slot within the \verb"dweak"-tabbed panel
of \verb"Physics" $\Rightarrow$ \verb"Subdomain Settings ...",
and is given as
\begin{equation}
\label{eq:COMSOL_dweak}
\begin{tabular}{l}
\verb"cbar2*x*(Haxitt*test(Haxi)"\\
\verb"+Hazitt*test(Hazi)"\\
\verb"+Hradtt*test(Hrad))",
\end{tabular}
\end{equation}
where \verb"Haxitt" denotes the double partial derivative of \verb"Haxi" with respect to time, \emph{etc.}
The remaining 2nd and 3rd slots of the \verb"dweak"-tabbed panel should be zero-filled.
\subsection{Boundary conditions}
Here the constraints stated in subsection \ref{subsec:AxisymmetricBCs}
are expressed in COMSOL-ready forms.
The model resonator's boundary conditions are defined through the
\verb"Physics" $\Rightarrow$ \verb"Boundary Settings ..." control panel.
On the left of this panel, select the \verb"Groups" tab. Each named boundary
\verb"Group" here corresponds to a particular electromagnetic boundary condition,
the most essential of which are described here. These different e.m.~boundary
conditions are specified by the expressions that populate the
three slots within their respective \verb"contr"-tabbed sub-panels, located
on the right-hand side of \verb"Boundary Settings ..."; `\verb"contr"' here
stands for `constraint'. [The neighboring \verb"weak"-tabbed and
\verb"dweak"-tabbed panels within the \verb"Boundary Settings ..." need not be touched
(and left zero-filled).]
\subsubsection{Electric wall (for a bounded isotropic dielectric)}
\begin{equation}
\label{eq:COMSOLExplWkFrmEWS1}
\begin{tabular}{l}
\verb"Hrad*nx+Haxi*ny";
\end{tabular}\\
\end{equation}
\begin{equation}
\label{eq:COMSOLExplWkFrmEWS4}
\begin{tabular}{l}
\verb"-Haxix+Hrady";
\end{tabular}\\
\end{equation}
\begin{equation}
\label{eq:COMSOLExplWkFrmEWS3}
\begin{tabular}{l}
\verb"(Hazi*nx-Hrad*M*nx"\\
\verb"-Haxi*M*ny+Hazix*nx*x"\\
~~~~~\verb"+Haziy*ny*x)/x";
\end{tabular}\\
\end{equation}
here \verb"nx" and \verb"ny" are, as `geometric variables' within COMSOL (in 2D),
the components of the (outward) unit normal vector on the boundary of a subdomain.
\subsubsection{Magnetic wall (for a bounded isotropic dielectric)}
\begin{equation}
\label{eq:COMSOLExplWkFrmMWS1}
\begin{tabular}{l}
\verb"Haxi*nx-Hrad*ny";
\end{tabular}\\
\end{equation}
\begin{equation}
\label{eq:COMSOLExplWkFrmMWS4}
\begin{tabular}{l}
\verb"Hazi";
\end{tabular}\\
\end{equation}
\begin{equation}
\label{eq:COMSOLExplWkFrmMWS3}
\begin{tabular}{l}
\verb"(Haxi*M*nx+Hazi*ny"\\
\verb"-Hrad*M*ny-Haziy*nx*x"\\
~~~~~\verb"+Hazix*ny*x)/x".
\end{tabular}\\
\end{equation}
\subsubsection{Radiation match (in free-space)}
As has already been discussed in subsection~\ref{subsec:AxisymmetricBCs},
the constraints appropriate to implementing a radiation match,
can be regarded (complex) linear combinations or `mixings' of pure
electric- and magnetic- wall constraints.
The first constraint mixes the magnetic-wall constraint~\ref{eq:magneticwallHcylcomp2},
\emph{i.e.}~\ref{eq:COMSOLExplWkFrmMWS4} above,
with the electric-wall constraint~\ref{eq:electricwallEcylcomp2isotrop},
\emph{i.e.}~\ref{eq:COMSOLExplWkFrmEWS3} above:
\begin{equation}
\label{eq:COMSOLExplWkFrmRMS1}
\begin{tabular}{l}
\verb"-i*cMW*Hazi*cbar*mf"\\
\verb"+cEW*(Hazi*nx-Hrad*M*nx"\\
\verb"-Haxi*M*ny+Hazix*nx*x"\\
\verb"+Haziy*ny*x)/x";
\end{tabular}\\
\end{equation}
note that `\verb"i"' here is the square root of minus one.
And the second constraint mixes
the electric-wall constraint~\ref{eq:electricwallEcylcomp1},
\emph{i.e.}~\ref{eq:COMSOLExplWkFrmEWS4} above,
with the magnetic-wall constraint~\ref{eq:magneticwallHcylcomp1},
\emph{i.e.}~\ref{eq:COMSOLExplWkFrmMWS1} above:
\begin{equation}
\label{eq:COMSOLExplWkFrmRMS2}
\begin{tabular}{l}
\verb"-i*cEW*(-Haxix+Hrady)"\\
\verb"+cMW*cbar*mf*(Haxi*nx-Hrad*ny)".
\end{tabular}\\
\end{equation}
Here, the pair of constants \{\verb"cMW" and \verb"cEW"\},
are defined through TABLE~\ref{tab:COMSOLConstants}.
When they are set to their standard (default) values
of \{ $1/\sqrt{2}$, $1/\sqrt{2}$ \}, equations
\ref{eq:COMSOLExplWkFrmRMS1} and \ref{eq:COMSOLExplWkFrmRMS2}
impose a radiation match on tangential field components
at the impedance of plane e.m.~waves in free-space.
Here also, \verb"cbar"$= \bar{c} \equiv 2 \pi / c$; and
\verb"mf" is the mode's (center) frequency; both need to be
defined within \verb"Options" $\Rightarrow$ \verb"Constants ...".
--as per their corresponding rows in TABLE~\ref{tab:COMSOLConstants};

The final (optional) constraint mixes
the electric wall constraint \ref{eq:electricwallHcylcomp},
\emph{i.e.}~\ref{eq:COMSOLExplWkFrmMWS1}
with the magnetic wall constraint~\ref{eq:magneticwallDcylcomp},
\emph{i.e.}~\ref{eq:COMSOLExplWkFrmMWS3}:
\begin{equation}
\label{eq:COMSOLExplWkFrmRMS3}
\begin{tabular}{l}
\verb"tngM*cbar*mf*(Hrad*nx+Haxi*ny)"\\
\verb"-tngE*(Haxi*M*nx+Hazi*ny"\\
\verb"-Hrad*M*ny-Haziy*nx*x"\\
\verb"+Hazix*ny*x)/x".
\end{tabular}\\
\end{equation}
Here; the constants \verb"tngM" and \verb"tngE",
are also defined through TABLE~\ref{tab:COMSOLConstants}
and thereupon \verb"Options" $\Rightarrow$ \verb"Constants ...".
Setting \{\verb"tngM", \verb"tngE"\} = \{1,0\}(\{1,0\})
constrains the magnetic (electric) field to be
wholly tangential on the impedance-matching plane
as is characteristic of electromagnetic traveling waves.
The default setting for this third constraint was
(arbitrarily) \{\verb"tngM", \verb"tngE"\} =  \{1,0\}).
[It is remarked here that the author sought to implement the
radiation-matching constraints more directly and elegantly
with time derivatives,
\emph{i.e.},~replacing \verb"2*pi*mf*Hazi" by \verb"Hazit", and
similarly for \verb"Hrad" and \verb"Haxi". But COMSOL did not
generate the intended frequency factor when interpreting them.
He thus resorted to entering the expressions as stated
in equations \ref{eq:COMSOLExplWkFrmRMS1} through
\ref{eq:COMSOLExplWkFrmRMS3}, requiring \verb"mf" to be
set, by hand, for each mode.]

\subsection{Geometry}
Each resonator's geometry needs to be either constructed within or imported into COMSOL. COMSOL's manual
provides instructions on how to implement both. Though simple geometries (\emph{e.g.}~a cylinder of solid dielectric material
inside a cylindrical metal can) can be quickly constructed by hand within COMSOL, the author found it advantageous to define
the sets of quadrilateral subdomains into which many axisymmetric dielectric resonators can be readily decomposed
using MATLAB scripts, where the script was run to generate the resonator's medial cross-section. The
key lines in these scripts were those of the form
\begin{verbatim}
q1 = poly2vert([[x1,y1];
[x2,y2];[x3,y3];[x4,y4]]);
\end{verbatim}
this particular line defines a quadrilateral, named \verb"q1", whose vertices have the \verb"x-y" coordinates:
\verb"[x1,y1], ..." \verb"[x4,y4]". These quadrilaterals could then be imported into COMSOL by entering
a list comprising their names into \verb"File" $\Rightarrow$ \verb"Import" $\Rightarrow$ \verb"Geometry Objects".
Each complete MATLAB script (also available from the author upon request) included, where known/relevant, the
cryogenic shrinkages of the resonator's constituent materials.

\subsection{Meshing}
If constructed out of quadrilaterals, the resonator's geometry can be meshed either
into sub-quadrilaterals using \verb+Mesh+ $\Rightarrow$ \verb"Map Mesh". Else, the geometry can always be meshed
into (psuedo-random) triangles using \verb+Mesh+ $\Rightarrow$ \verb"Initial Mesh" with (recommended) mesh
refinement over selected areas (covering the bright spots of WG modes).
Note that, for the geometry to be \verb"Map Mesh"-able, the vertices of its internal quadrilaterals should generally
all meet at `cross-roads', where, at each, four vertices belonging to four separate quadrilaterals all
meet at a point, as opposed to `T-junctions', where, at each, two vertices belonging to two separate
quadrilaterals meet at a point along the boundary edge of a third quadrilateral.
The reader is advised to consult the COMSOL manual
(chapter `Meshing', section `Generating Meshes', subsection `Creating Mapped Meshes in 2D') for a fuller (though a still
not wholly satisfactory) explanation of this rather quirky requirement.
The meshing density can be controlled by,
for selected edges, activating and entering
an appropriate integer in the `\verb"Contrained edge element distribution"' field within of the \verb"Boundary" tab of
the \verb"Map Mesh" control panel.
With regard to both \verb"Map Mesh"-ability and control over meshing density,
it can be advantageous (or plain necessary) to divide odd-shaped
subdomains (\emph{e.g.}~an `L'-shaped region covered throughout by a single
spatially uniform dielectric material) into several, more simply shaped
adjoining subdomains (\emph{e.g.}, in the case of dividing up the
L-shaped subdomain, two or even three rectangular subdomains).
Note that the geometry will not mesh if the allocations of elements along too many edges are (inconsistently)
specified; in other words, the edge element distribution must not be over constrained.

\subsection{Assignments}
\subsubsection{Interiors of subdomains}
The (either hand-made or imported) quadrilaterals composing the resonator's cross-sectional geometry
are assigned to one of the defined dielectric \verb"Groups" via the \verb"Subdomain Settings ..."
$\Rightarrow$ \verb"Subdomains" tab. Activating (\emph{i.e.}~ticking) the
\verb"Select by group" option here aids the verification of assignments.
\subsubsection{Edges of subdomains}The external edges of quadrilaterals should be set be obey one of the three above-described \verb"Group" boundary
conditions via the \verb"Physics" $\Rightarrow$ \verb"Subdomain Settings ..." $\Rightarrow$ \verb"Subdomains" tab.
COMSOL appears to be smart enough to recognize those edges that are external on its own accord and automatically `ghosts' (grays)
internal edges; the latter should not be assigned to any boundary \verb"Group" condition. Thus,
with the \verb"Select by group" feature activated, all the appropriate edges can be assigned to
the appropriate (usually electric-wall) boundary condition in a single selection of the \verb"Groups" drop-down menu.

\subsection{Solution}
In \verb"Solve" $\Rightarrow$ \verb"Solve Parameters":

(a)  set the selected \verb"Solver:" to `\verb"Eigenvalue:"';

and, with the \verb"General" tab selected,

(b) set the \verb"Desired number of eigenvalues:" to `10' --or whatever ones desires;

(c) set \verb"Search for eigenvalue around:" to `0' --or whatever;

(d) set \verb"Linear system solver:" to `\verb"Direct (SPOOLES)" (this is at least the author's starting
recommendation);

(e) set \verb"Matrix symmetry:" to \verb"Symmetric"'.

Having implemented all of the above, one should now be able to \verb"Solve" $\Rightarrow$ \verb"Solve Problem".

\subsection{Postprocessing}
COMSOL Multiphysics' standard documentation explains how to configure and use of the
\verb"Postprocessing" $\Rightarrow$ \verb"Plot Parameters" control panel. Only a few
specific pointers are supplied here:

[1] The center frequencies of solved resonances
can be viewed through the \verb"Solution to use" $\Rightarrow$ \verb"Eigenvalue:"  drop-down
menu in the \verb"General"-tabbed sub-panel of the \verb"Plot Parameters" control panel.

[2] To display the morphology and features of the solved eigenmodes,
the \verb"Expression:" slot within the \verb"Surface"-tabbed sub-panel of the same is filled
with either (i) some function of the solved field variables \{\verb"Hrad", \verb"Hazi", \verb"Haxi" \}
or (ii) one of those expressions (\emph{e.g.}~\verb"ElecMagSqrd") pre-defined in \verb"Options" $\Rightarrow$
\verb"Expressions" \verb"Scalar Expressions" as COMSOL's equivalent of
TABLE~\ref{tab:COMSOLscalarexpress}, or (iii) some function/combination thereof. For example,
\begin{equation}
\label{eq:COMSOLPostProcSurf}
\begin{tabular}{l}
\verb"log10(AbsMagEnDens+10^(-10))"
\end{tabular}\\
\end{equation}
can be inserted to view the magnetic energy density on a logarithmic scale.
To view (as a diagnostic) the divergence of the magnetic field strength, which
should be zero, one inserts \verb"DivH" $\equiv$ \verb"(Hrad-Hazi*M+(Haxiy+Hradx)*x)/x"
instead.

[3] Determinations of an electromagnetic mode's volume, filling factor(s), and length ($\Lambda$),
as per equations \ref{eq:mode_volume}, \ref{eq:filling_factor}, and \ref{eq:length}, respectively,
all make use of the \verb"Postprocessing" $\Rightarrow$ \verb"Domain Integration" panel.
For example, the numerator
$\int \int \int_{\rm{h.-s.}} \epsilon |\textbf{E}|^2 \textrm{d} \textrm{V} $
on the right-hand-side of \ref{eq:mode_volume} can be
evaluated by inserting \verb"ElecEnDens" $\equiv$ \verb"Erad*Drad+Eazi*Dazi+Eaxi*Daxi"
into this panel's \verb"Expression:" slot, with those entries selected in the
\verb"Subdomain selection" list on the left-hand side of the same panel
covering all significant parts of the mode's bright spot.

[4] With regards to determining filling factors, the numerator on the right-hand side of
equation \ref{eq:filling_factor} can be evaluated by selecting only those subdomains
filled with the relevant dielectric (as opposed to free-space).

[5] To determine the resistive-wall-loss integral
$\int |\textbf{n} \pmb{\times} \textbf{H}|^2 \textrm{d} \textrm{S} $
(forming the denominator of equation \ref{eq:length}),
one uses the \verb"Postprocessing" $\Rightarrow$
\verb"Boundary Integration" panel
with \verb"2*pi*x*(abs(Hazi)^2" \verb"+abs((Haxi*nx" \verb"-Hrad*ny))^2)"
inserted into the \verb"Expression:" slot within
the \verb"Expression to integrate" box therein, and
where entries selected with \verb"Boundary selection"
correspond to the resonator's enclosing (metallic and lossy)
surfaces.

[6] The maximum/minimum of a field variable, as required to evaluate
the denominator on the right-hand side of \ref{eq:mode_volume},
can be determined through the \verb"Postprocessing" $\Rightarrow$ \verb"Plot parameters"
$\Rightarrow$ \verb"Max/Min", wherein the \verb"Expression:" slot is filled with the
field variable's expression (\emph{viz.}~\verb"ElecEnDens" for evaluating said denominator).

\end{document}